\documentclass[aps,  amssymb, amsmath, nobibnotes, prb, superscriptaddress]{revtex4-2}
\usepackage{amsmath}
\usepackage{amssymb}
\usepackage{graphicx}
\usepackage{bm}
\usepackage{amsfonts}
\usepackage[T1]{fontenc}
\usepackage[bookmarks=false,linkcolor=blue,urlcolor=blue,colorlinks,citecolor=blue]{hyperref}

\begin{document}

\title{Coulomb drag by motion of a monolayer polar crystal through  graphene nanoconstriction}
\author{A. L. Chudnovskiy}
\affiliation{I. Institut f\"ur Theoretische Physik, Universit\"at Hamburg, Notkestra\ss e 9, D-22607 Hamburg, Germany}

\begin{abstract} 
	We theoretically predict that the motion of a polar crystalline layer between two graphene planes exerts Coulomb drag on electrons in graphene, inducing a DC drag current. The physical mechanism underlying this drag arises from inter-valley scattering of charge carriers in graphene caused by the time-dependent potential of the moving crystalline layer.
	This drag effect manifests above a finite threshold doping of the graphene layers, which is determined by the lattice structure and the relative orientation of the crystalline layer with respect to the graphene lattice. Additionally, the drag current exhibits a nonlinear Hall effect due to the interplay between the induced nonequilibrium pseudospin and the intrinsic pseudospin-orbit coupling in graphene.
	In turn, the drag exerted on electrons in graphene produces a back-action on the crystalline layer, which can be described as an increase in its dynamic viscosity.
\end{abstract}

	\maketitle
	
\section{Introduction}	

Water transport through nano-porous devices has demonstrated significant potential as an effective technology for water desalination, energy conversion, and molecular sensing \cite{GrapheneMembranes_2024,GUNAY2023114785,NanoConfinedChannels2019}. Rapid experimental advancements in the fabrication of artificial nano-membranes based on carbon nanotubes and graphene nano-constrictions highlight the need for a theoretical understanding of both the principles governing water flow through nano-constrictions and the physical processes induced in the walls of a constriction by the water flow. 

 In particular,  recent experiments have shown that confining water in nano-constrictions leads to an ordering phase transition, resulting in a lattice of water molecules and forming a two-dimensional ice crystal \cite{SquareIce_Geim2015,Trushin2022,Sahin2018,li2019twodimensional}. Due to the inherent dipole moment of a water molecule, this single-layer ice forms a polar crystal with a frozen pattern of dipoles. Under applied external pressure, the ice layer can move through constrictions. A recent example of such motion is demonstrated in experiments in Refs. \cite{Geim2019, Geim2012,FastWaterTransport,UnderstandingWaterTransport}, where water flowed through a nano-constriction of a single graphene layer's width between graphene planes. The motion of the spatially ordered dipole moments of the ice layer generates a time- and space-periodic electric field acting on the electrons in the walls of the constriction. Notably, the spatial period of the electric field set by the lattice spacing of the polar crystal  is comparable to the nearest neighbor distance in graphene. For instance, for the crystal of square ice, the lattice spacing is given by $a_0\approx 2.8$ \AA \,  \cite{SquareIce_Geim2015}, while the nearest neighbor distance in graphene  is $a=1.42$ \AA \, \cite{KatsnelsonBook2012}.  Consequently, the momentum transferred from the moving crystal to the electrons in graphene is close to the distance between the $K$ and $K'$ points in reciprocal space. This opens the possibility for Coulomb drag of electrons in graphene induced by the motion of the polar crystalline layer, facilitated by inter-valley and/or umklapp scattering processes.

In this paper, we investigate the DC-drag current density in graphene induced by the motion of a crystalline layer. We uncover several distinct features of the DC-drag current that set it apart from conventional drag phenomena:
(i) The drag current is proportional to the velocity of the crystalline layer.
(ii) The direction of the drag current is determined by the orientation of the moving crystalline lattice with respect to the vector connecting the two K-points in graphene, rather than the direction of the crystalline layer's velocity.
(iii) The drag exhibits signatures of the nonlinear Hall effect, which can be traced back to the excitation of nonequilibrium pseudospin dipole by the potential of the moving crystalline layer.  
(iv) The drag sets in at a finite doping level that creates Fermi surfaces around the K points.
(v) Unlike conventional Coulomb or phonon drag, the leading contribution to the drag current is temperature-independent.

\begin{figure}[t!]
	\centering
	\includegraphics[width=\textwidth]{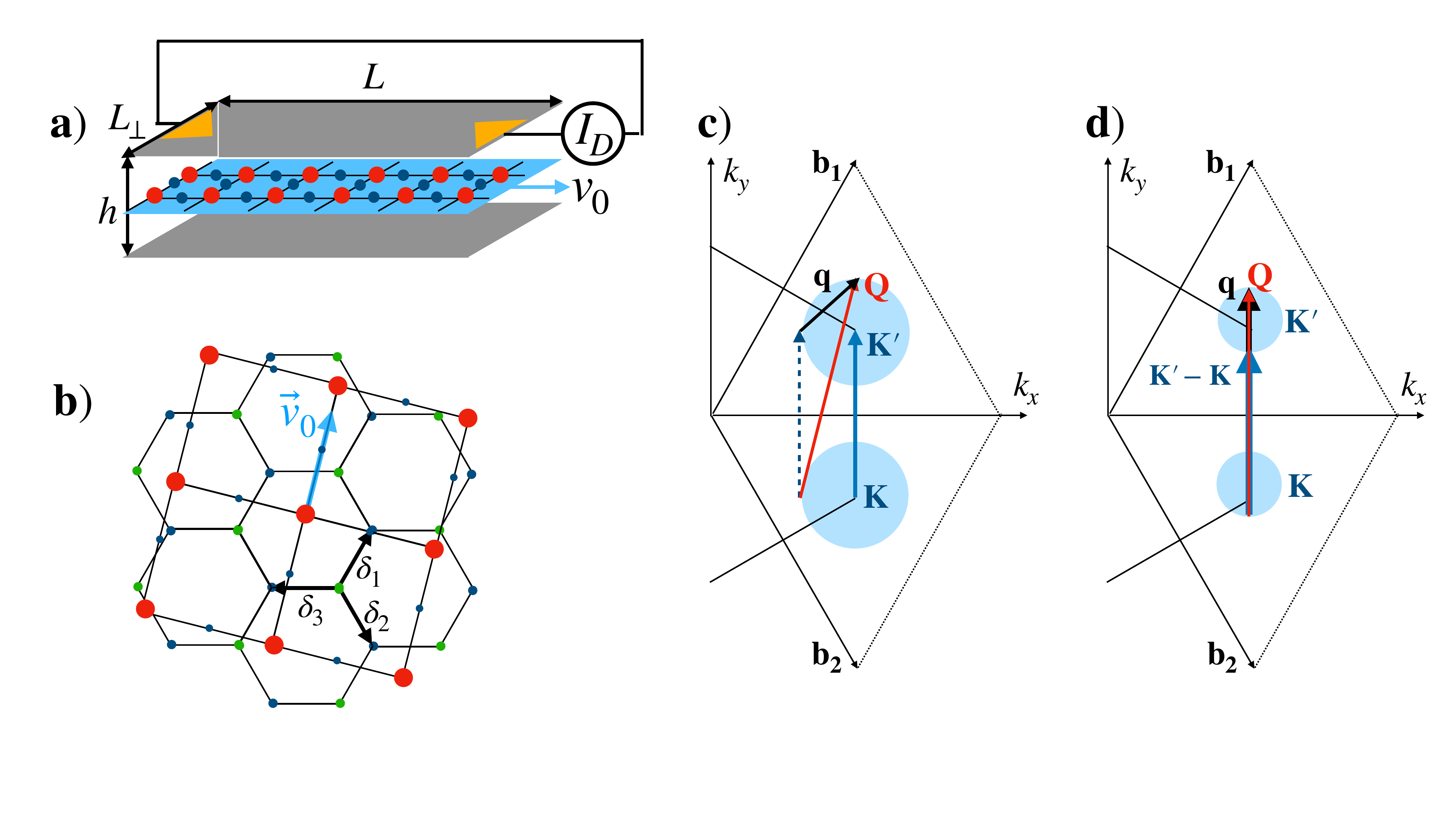}
	\vskip -.3cm
	\caption{ a) Scheme of motion of a crystalline layer through a graphene nano-constriction.  The negative (red) and positive (blue) charges ions form a frozen pattern of in-plane dipole moments. The drag current can be measured in one of the layers forming the constriction.  b) Graphene and square ice lattices. The maximal energy transfer  $\omega=({\bf v}_0\cdot {\bf Q} )$ is achieved when the velocity of the ice ${\bf v}_0$ is parallel to one of the main lattice directions  (see the text). c) Reciprocal lattice of graphene. The vectors $\bf b_1$, $\bf b_2$ indicate the first Brillouin zone.  Filled circles denote the Fermi sea around the $K$ and $K'$ points.  Arrows indicate the momentum conservation by inter-valley transitions leading to the drag current ${\bf K'}-{\bf K}+{\bf q}={\bf Q}$.  d) Momentum conservation  at the drag threshold for ${\bf Q}\parallel {\bf K'}-{\bf K}$,  $|{\bf q}|=2p_F$.}
	\label{fig:Transitions}
\end{figure}

The scattering mechanism underlying the drag effect is similar to the umklapp scattering that causes excess resistivity in graphene/hexagonal boron nitride heterostructures \cite{Umklapp_2019}. However, unlike the stationary superlattice in heterostructures, scattering by the moving crystalline layer transfers energy to electrons in graphene, thereby inducing a finite drag current.
The drag effect considered here also bears some analogy to the acousto-electric effect, where electric current is generated by acoustic waves \cite{Parmenter53, Eckstein64, Many1969, Falko93, ValleyDrag2019, Lapa2020}. The main difference lies in the very small wavelength associated with the potential of the moving polar crystal, which can induce inter-valley transitions or umklapp processes.
Although motivated by recent experiments on the ice layer sliding through the nano-constriction, the drag phenomenon is of a quite general nature and is applicable to any two-dimensional (2D) polar crystal layer   (see Fig. \ref{fig:Transitions}a).  

\section{Model of interaction between the crystalline layer and graphene}
Let us introduce the model for the Coulomb drag by a moving polar crystalline layer.  The crystalline layer exerts a periodic potential on electrons in the immediate vicinity determined by its crystal structure, given by $U({\bf r})\sim \sum_{i=1,2}\cos\left({\bf Q}_i {\bf r}\right)$,   where ${\bf Q}_{1,2}$ 
denote the reciprocal lattice vectors of the crystal. The potential of the moving crystal gains time dependence through a boost transformation ${\bf r}\rightarrow {\bf r}-{\bf v}_0 t$, where ${\bf v}_0$  denotes the velocity of the crystalline layer. Therefore, we  model the periodic time-dependent potential exerted by the layer moving with velocity $v_0$ by the expression  
$
	U({\bf r, t})=\sum_{i=1,2}U_{i}\cos({\bf Q}_{i}({\bf r}-{\bf v}_0 t)). 
$
 The scalar products of the  vectors  ${\bf Q}_{1,2}$ with the velocity of the crystalline layer determine the frequencies of the time dependent potential $\omega_{i}=({\bf Q}_{i} \cdot {\bf v}_0)$. As will be shown below, the amplitude of the drag current is proportional to the frequency of the driving potential, hence the component of the potential with the higher frequency provides the dominant contribution to the drag. 
  For instance, among several possible orderings of water molecules,  the simplest one is a square ice, in which molecules are ordered in a one-layer square lattice.  In this configuration  the vectors ${\bf Q}_1$ and ${\bf Q}_2$ are orthogonal.  The maximal drag is achieved when  one reciprocal wave vector, ${\bf Q}_1$, is aligned with the velocity of the layer ${\bf v}_0$. In this case, the other wave vector ${\bf Q}_2$ is perpendicular to the velocity, and its component provides no contribution to the drag. Therefore, in what follows we consider only one component of the periodic potential and drop the subscript at the wave vector ${\bf Q}$.  
 The external potential of the crystalline layer affects both the on-site energies and the nearest-neighbor hopping amplitudes in the graphene lattice. Let us consider these two effects separately. 
 
\subsection{Modulation of on-site energies}

We introduce the basis wave functions  $\psi_{A,B}({\bf r})$ localized at the positions of the atoms of the two ($A,B$) graphene sublattices  \cite{RevModPhys.81.109, KatsnelsonBook2012}. Thereby the atoms of the $A$-sublattice are situated at ${\bf r}_{nm}=n{\bf a}_1+m{\bf a}_2$, where $n,m$ are integers and ${\bf a}_{1,2}$ denote the lattice vectors in graphene, whereas the positions of  atoms of the $B$-sublattice are given by ${\bf r}_{nm}+{\bm \delta}_3$ (see Fig.  \ref{fig:Transitions}b).  The action of the periodic potential on the on-site energies  of the atoms is given by 
\begin{eqnarray}
	\nonumber && 
	H_{U}=U_0 \sum_{n,m}\left\{\cos({\bf Q}{\bf r}_{nm}-\omega t)\bar{\psi}_A({\bf r}_{nm}) \psi_A({\bf r}_{nm})+
	\cos({\bf Q}({\bf r}_{nm}+{\bm \delta}_3)-\omega t)\bar{\psi}_B({\bf r}_{nm}+{\bm\delta}_3) \psi_B({\bf r}_{nm}+{\bm\delta}_3)\right\}.
\end{eqnarray}
Performing Fourier-transformation according to
\begin{equation}
	\psi_A({\bf r})=\int_{\bf k} \psi_A({\bf k}) e^{i{\bf k}{\bf r}}, \, \, \, \psi_B({\bf r}+{\bm \delta}_3)=\int_{\bf k} \psi_B({\bf k}) e^{i{\bf k}{\bf r}}, 
	\label{FT}
\end{equation}
we cast the external potential to the following expression 
\begin{eqnarray}
	\nonumber && 
	H_{U}=\frac{U_0}{2}  e^{-i\omega t} e^{\frac{i}{2}{\bf Q} {\bm\delta}_3} \int_{\bf k} \left\{ \cos\left(\frac{{\bf Q}}{2}{\bm\delta}_3\right) \left[\bar{\psi}_A({\bf k}+{\bf Q})\psi_A({\bf k})+\bar{\psi}_B({\bf k}+{\bf Q})\psi_B({\bf k})\right] \right. \\ 
	\nonumber && \left. 
	-i \sin\left(\frac{{\bf Q}}{2}{\bm\delta}_3\right) \left[\bar{\psi}_A({\bf k}+{\bf Q})\psi_A({\bf k})-\bar{\psi}_B({\bf k}+{\bf Q})\psi_B({\bf k})\right]\right\}+ \\ 
	\nonumber && 
	e^{i\omega t} e^{-\frac{i}{2}{\bf Q} {\bm\delta}_3} \int_{\bf k} \left\{ \cos\left(\frac{{\bf Q}}{2}{\bm\delta}_3\right) \left[\bar{\psi}_A({\bf k}-{\bf Q})\psi_A({\bf k})+\bar{\psi}_B({\bf k}-{\bf Q})\psi_B({\bf k})\right] \right. \\ 
	&& \left. 
	+i \sin\left(\frac{{\bf Q}}{2}{\bm\delta}_3\right) \left[\bar{\psi}_A({\bf k}-{\bf Q})\psi_A({\bf k})-\bar{\psi}_B({\bf k}-{\bf Q})\psi_B({\bf k})\right]\right\}. 
\end{eqnarray}
Furthermore, using the pseudospin spinors 
$ \Psi({\bf k})=(\psi_A({\bf k}), \psi_B({\bf k}))^T$, the Hamiltonian $H_{U}$ can be written as 
\begin{eqnarray}
	\nonumber && 
	H_{U}=\frac{U_0}{2}  e^{-i\omega t} e^{\frac{i}{2}{\bf Q} {\bm\delta}_3} \int_{\bf k} \left\{ \cos\left(\frac{{\bf Q}}{2}{\bm\delta}_3\right) \bar{\Psi}({\bf k}+{\bf Q})\sigma_0\Psi({\bf k})
	-i \sin\left(\frac{{\bf Q}}{2}{\bm\delta}_3\right) \bar{\Psi}({\bf k}+{\bf Q})\sigma_z\Psi({\bf k})\right\}+ \\ 
	&& 
		\frac{U_0}{2}   e^{i\omega t} e^{-\frac{i}{2}{\bf Q} {\bm\delta}_3} \int_{\bf k} \left\{ \cos\left(\frac{{\bf Q}}{2}{\bm\delta}_3\right) \bar{\Psi}({\bf k}-{\bf Q})\sigma_0\Psi({\bf k})
	+i \sin\left(\frac{{\bf Q}}{2}{\bm\delta}_3\right) \bar{\Psi}({\bf k}-{\bf Q})\sigma_z\Psi({\bf k})\right\},  
	\label{H0z}
\end{eqnarray}
where $\sigma_0={\bf 1}_2$ and $\sigma_z$ denotes the Pauli matrix. Assuming ${\bf Q}$ to be close to ${\bf K'}-{\bf K}$, we introduce the parametrization ${\bf Q}={\bf K'}-{\bf K}+{\bf q}$.   The external potential induces transitions between the electron states in different valleys.    Then, assuming $|q|\ll |Q|$,  we can represent Eq. (\ref{H0z}) as 
\begin{eqnarray}
	\nonumber && 
	H_{U}=\frac{U_0}{2}  e^{-i\omega t} e^{\frac{i}{2}({\bf K'} -{\bf K}+{\bf q}){\bm\delta}_3} \int_{\bf p} \left\{ \cos\left(\frac{{\bf K'} -{\bf K}+{\bf q}}{2}{\bm\delta}_3\right) \bar{\Psi}_{K'}({\bf p}+{\bf q})\sigma_0\Psi_{K}({\bf p})
	-i \sin\left(\frac{{\bf K'} -{\bf K}+{\bf q}}{2}{\bm\delta}_3\right) \bar{\Psi}_{K'}({\bf p}+{\bf q})\sigma_z\Psi_{K}({\bf p})\right\} \\ 
	&& 
	+ 	\frac{U_0}{2}   e^{i\omega t} e^{-\frac{i}{2}({\bf K'} -{\bf K}+{\bf q}) {\bm\delta}_3} \int_{\bf p} \left\{ \cos\left(\frac{{\bf K'} -{\bf K}+{\bf q}}{2}{\bm\delta}_3\right) \bar{\Psi}_{K}({\bf p})\sigma_0\Psi_{K'}({\bf p}+{\bf q})
	+i \sin\left(\frac{{\bf K'} -{\bf K}+{\bf q}}{2}{\bm\delta}_3\right) \bar{\Psi}_{K}({\bf p})\sigma_z\Psi_{K'}({\bf p}+{\bf q})\right\}.  
	\label{Hz}
\end{eqnarray}
Further simplification is reached by noticing that $({\bf K'}-{\bf K})\cdot {\bm\delta}_3=0$, which results in 
\begin{eqnarray}
	\nonumber && 
	H_{U}=\frac{U_0}{2}  e^{-i\omega t} e^{\frac{i}{2}{\bf q}{\bm\delta}_3} \int_{\bf p} \left\{ \cos\left(\frac{{\bf q}}{2}{\bm\delta}_3\right) \bar{\Psi}_{K'}({\bf p}+{\bf q})\sigma_0\Psi_{K}({\bf p})
	-i \sin\left(\frac{{\bf q}}{2}{\bm\delta}_3\right) \bar{\Psi}^{K'}({\bf p}+{\bf q})\sigma_z\Psi_{K}({\bf p})\right\}+ \\ 
	&& 
	\frac{U_0}{2}  e^{i\omega t} e^{-\frac{i}{2}{\bf q} {\bm\delta}_3} \int_{\bf p} \left\{ \cos\left(\frac{{\bf q}}{2}{\bm\delta}_3\right) \bar{\Psi}_{K}({\bf p})\sigma_0\Psi_{K'}({\bf p}+{\bf q})
	+i \sin\left(\frac{{\bf q}}{2}{\bm\delta}_3\right) \bar{\Psi}_{K}({\bf p})\sigma_z\Psi_{K'}({\bf p}+{\bf q})\right\}.  
	\label{Hzq}
\end{eqnarray}

\subsection{Change of the hopping amplitude by the potential of the moving lattice}

We assume  that the change of the hopping amplitude between the nearest neighbor atoms is proportional to the local external potential in the middle of the link connecting the two atoms in the graphene lattice. This assumption leads to the following Hamiltonian for the change of the nearest-neighbor hopping 
\begin{equation}
	H_{h}=U_1 \sum_{ {\bm\delta}_1,{\bm\delta}_2,{\bm\delta}_3} \sum_{n,m}\left\{\cos\left[{\bf Q}\left({\bf r}_{nm}+\frac{{\bm\delta}_{\nu}}{2}\right)-\omega t\right]\left[\bar{\psi}_A({\bf r}_{nm}) \psi_B({\bf r}_{nm}+{\bm\delta}_{\nu})+
	\bar{\psi}_B({\bf r}_{nm}+{\bm\delta}_\nu) \psi_A({\bf r}_{nm})\right]\right\}
	\label{Hh}
\end{equation}
Here the coupling constant $U_1$ is in general different from the coupling $U_0$ that describes the change of the local on-site potential. Performing  the Fourier transformation in Eq. (\ref{Hh}) into the ${\bf k}$ space and leaving only the states close to  $K$ and $K'$ point, we cast the Fourier transformed Hamiltonian $H_{h}$ to the form 
\begin{eqnarray}
	\nonumber && 
	H_{h}=\frac{U_1}{2}\sum_{i=1}^3\int_{\bf p} \left\{ e^{-i\omega t} e^{\frac{i}{2}{\bm \delta}_3({\bf K'}-{\bf K}+{\bf q})} 
	\bar{\Psi}_{K'}({\bf q}+{\bf p}) \left(\sigma_x\cos\left[\left(\frac{{\bf K}+{\bf K'}+{\bf q}}{2}+{\bf p}\right)({\bm \delta}_i-{\bm \delta}_3)\right]-
	\right. \right. \\ 
	\nonumber && 
	\left. 
	\sigma_y\sin\left[\left(\frac{{\bf K}+{\bf K'}+{\bf q}}{2}+{\bf p}\right)({\bm \delta}_i-{\bm \delta}_3)\right]\right)\Psi_{K}({\bf p}) +  \\ 
	\nonumber && 
	e^{i\omega t} e^{-\frac{i}{2}{\bm \delta}_3({\bf K'}-{\bf K}+{\bf q})} 
	\bar{\Psi}_{K}({\bf p}) \left(\sigma_x\cos\left[\left(\frac{{\bf K}+{\bf K'}+{\bf q}}{2}+{\bf p}\right)({\bm \delta}_i-{\bm \delta}_3)\right]- \right. \\ 
	&& \left. \left.
	\sigma_y\sin\left[\left(\frac{{\bf K}+{\bf K'}+{\bf q}}{2}+{\bf p}\right)({\bm \delta}_i-{\bm \delta}_3)\right]\right)\Psi_{K'}({\bf p}+{\bf q})\right\}.  
	\label{Hxy}
\end{eqnarray}

\subsection{Total drag Hamiltonian}

For the states close to the $K$ and $K'$ points, the electron spectrum  in the graphene layer given by a well-known Hamiltonian \cite{RevModPhys.81.109,KatsnelsonBook2012} 
\begin{equation}
	H_g=-iv_F\! \int d^2{\bf r}\left[\Psi_1^+({\bf r}){\bm \sigma}\cdot\! \nabla \Psi_1({\bf r})+ \Psi_2^+({\bf r}){\bm \sigma^*}\cdot\! \nabla \Psi_2({\bf r})\right],
	\label{graphene}
\end{equation}
where the indexes $1,2$ correspond to the $K$ and $K'$ points respectively, and $\Psi({\bf k})=(\psi_{A}({\bf k}), \psi_{B}({\bf k}))^T$ denotes a spinor in the pseudospin space. Because the electron spin does not play a role in the drag physics considered here, spin indexes are suppressed. The interaction with the moving crystalline layer is given by Eqs. (\ref{Hzq}), (\ref{Hxy}).  Upon projecting  on the states close to the $K$ points, the drag potential acquires the form 
\begin{equation}
	H_d=
	\frac{U_0}{2}\int_{\bf p}  \left\{ \Psi^+_{1}({\bf p}+{\bf q})\left({\bf u}({\bf q}, {\bf p})\cdot  {\bm \sigma}\right) \Psi_2({\bf p}) e^{-i\omega t} +
	\Psi^+_2({\bf p}) \left({\bf u}^*({\bf q}, {\bf p})\cdot {\bm\sigma}\right)   \Psi_1({\bf p}+{\bf q}) e^{i\omega t}\right\}, 
	\label{drag_RWA}
\end{equation}
where $\left({\bf u}({\bf q}, {\bf p})\cdot  {\bm \sigma}\right) =\sum_{i=0}^3 u_i({\bf q}, {\bf p}) \sigma_i$, and we set $U_0$ to be the energy scale characterizing the interaction strength. 
In writing Eq. (\ref{drag_RWA}) we retained only the time-dependent exponentials that can satisfy the energy conservation condition $v|{\bf p}|\pm \omega=v|{\bf p}\pm {\bf q}|$ according to the dispersion in graphene close to the $K$-points. 

The components of the vector of coupling constants ${\bf u}$ are obtained by comparison with Eqs. (\ref{Hzq}), (\ref{Hxy}) as follows 
\begin{eqnarray}
	&& u_0= \frac{1}{2} e^{\frac{i}{2}{\bf q}{\bm\delta}_3} \cos\left(\frac{{\bf q}}{2}{\bm\delta}_3\right), \label{u0} \\ 
	&& u_3= -i \frac{1}{2} e^{\frac{i}{2}{\bf q}{\bm\delta}_3}  \sin\left(\frac{{\bf q}}{2}{\bm\delta}_3\right), \label{u3} \\ 
	&& u_1= \frac{U_1}{2U_0} e^{\frac{i}{2}{\bf q}{\bm \delta}_3}  \sum_{i=1}^3 \cos\left[\left(\frac{{\bf K}+{\bf K'}+{\bf q}}{2}+{\bf p}\right)({\bm \delta}_i-{\bm \delta}_3)\right] =\frac{U_1}{2U_0} e^{\frac{i}{2}{\bf q}{\bm \delta}_3} \left(1- \sum_{i=1}^2\cos\left[\left(\frac{{\bf q}}{2}+{\bf p}\right)({\bm \delta}_i-{\bm \delta}_3)\right] \right), \label{u1} \\ 
	&& u_2=-\frac{U_1}{2U_0} e^{\frac{i}{2}{\bf q}{\bm \delta}_3}  \sum_{i=1}^3 \sin\left[\left(\frac{{\bf K}+{\bf K'}+{\bf q}}{2}+{\bf p}\right)({\bm \delta}_i-{\bm \delta}_3)\right] = \frac{U_1}{2U_0} e^{\frac{i}{2}{\bf q}{\bm \delta}_3}  \sum_{i=1}^2 \sin\left[\left(\frac{{\bf q}}{2}+{\bf p}\right)({\bm \delta}_i-{\bm \delta}_3)\right], \label{u2}
\end{eqnarray}
where in the second equalities we took into account $\frac{{\bf K}+{\bf K'}}{2}\cdot({\bm \delta}_1-{\bm \delta}_3)=\frac{{\bf K}+{\bf K'}}{2}\cdot({\bm \delta}_2-{\bm \delta}_3)=\pi$. Here we use the coupling $U_0$ as the overall scale for the coupling strengths. 

\section{External potential as a source of inter-valley and umklapp scattering}

Consider now the action of external potential  Eqs. (\ref{drag_RWA}) as a source of the inter-valley and umklapp scattering.  An elementary scattering process requires conservation of energy and quasi-momentum (with $\hbar=1$ throughout this paper) 
\begin{equation}
\epsilon({\bf p'})=\epsilon({\bf p})+\omega , \, \, \,	{\bf p}+{\bf Q}=	{\bf p'}+n_1{\bf b}_1+n_2{\bf b}_2+m({\bf K}-{\bf K'}). 
	\label{shell-conditions}
\end{equation} 
Here $n_1$, $n_2$, and $m$ are integers, ${\bf b}_1$ and ${\bf b}_2$ denote the reciprocal lattice vectors in graphene, and ${\bf K}$ and ${\bf K'}$ denote the wave vectors at $K$-points. The wave vectors ${\bf p}$ and ${\bf p'}$ relate to the initial and the final states by the scattering event.  The relation of wave-vectors for a scattering event with $n_1=n_2=0$, $m=1$ is shown in Fig. \ref{fig:Transitions}b). Using the data on the graphene lattice structure one can conclude that the minimal mismatch of the vectors ${\bf p}$ and 	${\bf p'}$ is reached by the inter-valley scattering, when the reciprocal lattice vector ${\bf Q}$ of the crystalline layer is parallel to the vector ${\bf K'}-{\bf K}$ connecting the different $K$-points. Taking the length of the vector ${\bf Q}$ proper to the square ice lattice, $| {\bf Q}|=2\pi/a_0$, using the standard values for the $K$ points 
\begin{equation}
	{\bf K}=\frac{2\pi}{3a} (1, 1/\sqrt{3}), \, \, \, {\bf K'}=\frac{2\pi}{3a} (1, -1/\sqrt{3}), 
	\label{K-points}
\end{equation}
and considering the geometry ${\bf Q}\parallel {\bf K}-{\bf K'}\parallel {\bf \hat y}$,  we  estimate the difference of the final and initial wave-vectors as 
\begin{equation}
	{\bf q}= {\bf p'}-{\bf p}={\bf Q}- ({\bf K'}-{\bf K})\approx   \frac{0.77 }{a} {\bf \hat y}.
	\label{q-mismatch}
\end{equation}
Absorption of the momentum $ {\bf q}$ by the electron system of graphene results in the drag current.  Furthermore, energy conservation by scattering requires that  the vector ${\bf q}$ connects two states in the graphene energy band, that is $\epsilon({\bf p}+{\bf q})=\epsilon({\bf p})+\omega$. 
The maximal possible value of the frequency $\omega$ is given by $\omega=v_0 |Q|$. However, since the drift velocity of the ice layer $v_0\sim 1$ m/s  \cite{Geim2019} is much smaller that the electron velocity in graphene $v_F\sim 10^6$ m/s, the energy of the transition between the states with ${\bf p}$  parallel to ${\bf p}+{\bf q}$, given by $v_F |q|$, is much larger than $\omega$. The energy conservation condition can be satisfied though, if the wave-vectors of the initial and final states are anti-parallel, lying at opposite points of the Fermi-surface, as shown in Fig. \ref{fig:Transitions}c). Then the energy conservation reads $v_F(|q|-2p_F)=v_0 |Q|$, and it can be satisfied at the Fermi wave vector 
$p_F=\frac{1}{2}(|q|-\frac{v_0}{v_F}|Q|)$. This Fermi wave vector marks the threshold for the Fermi energy $\mu= v_F  p_F$, at which the drag current starts. For larger Fermi wave vectors, the conservation of energy and quasi-momentum can also be satisfied at some angle between the vectors ${\bf p}$ and ${\bf q}$, as shown in Fig. \ref{fig:Transitions}d). Therefore, the inter-valley scattering provides the most efficient drag mechanism under the conditions where the velocity and one crystallographic axis of the moving crystalline layer are parallel to the ${\bf K}-{\bf K'}$ vector in graphene. 

Reaching the threshold chemical potential for the drag requires substantial doping, which nevertheless lies within the reach of modern experiments. Using the values $v_F\sim 10^6$m/s, $v_0\sim 1$ m/s, and $|{\bf q}|\sim |{\bf Q}-({\bf K'}-{\bf K})|\sim 5.5 \cdot 10^9$m$^{-1}$ (see Eq. \ref{q-mismatch}), we  estimate the threshold chemical potential as $\mu \sim 0.6 t \sim 2$ eV. It is noteworthy that the threshold chemical potential is still below the van Hove singularity at the M-point of graphene $E_{\mathrm{vH}}\sim 3 eV$, so considering separate $K$ points remains valid. In light of recent progress in doping graphene up to and even above the M-point   \cite{Doping_PhysRevB.100.035445,Doping_PhysRevLett.125.176403}, reaching the threshold doping for the drag looks quite feasible. We estimate the electron density around the threshold as $n_e\sim 10^{18}$ m$^{-2}$. 

At the same time, at such a high doping, taking account the deviation of the electron dispersion from the linear one adopted in Eq. (\ref{graphene}) may be necessary for the close comparison with experiment. We do not believe however, that the deviation from the linear dispersion would lead to qualitative changes of results, so we will use the approximation of the linear dispersion for the proof of principle.

\section{Calculation of the drag current}
We  calculate the drag current in a single graphene layer forming the constriction in frame of perturbative expansion in the interaction strength $U_0$ using the Keldysh formalism \cite{DiVentra_2008,kamenev2023field,Sieberer_2016,Keldysh_FRG_PhysRevB.95.125412,Aoki_PhysRevB.79.081406}.  The total current is then given by the sum of the currents in the two (the upper and the lower) layers. 

\begin{figure}[t!]
	\centering
	\includegraphics[width=0.5\textwidth]{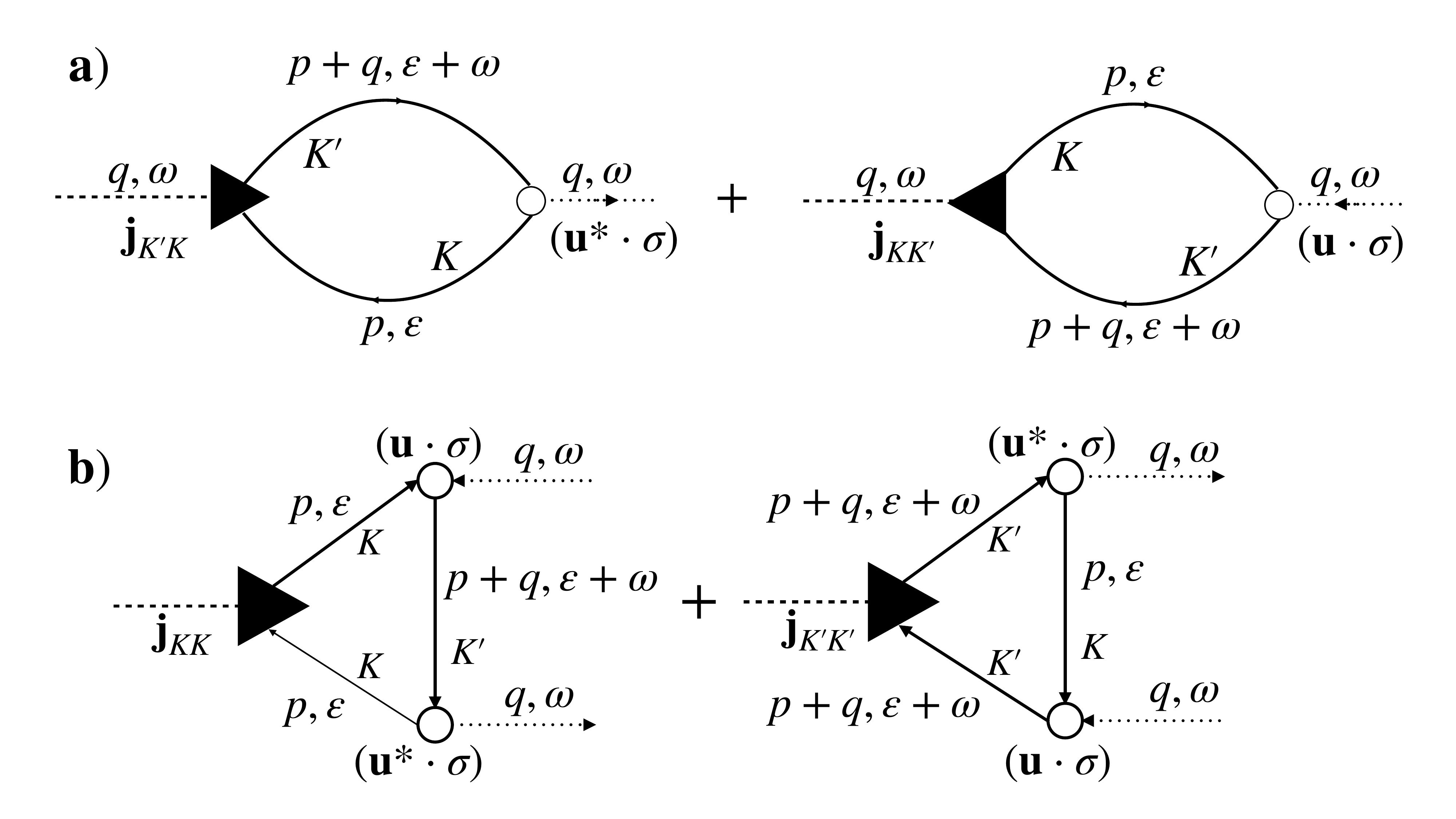}
	\caption{Diagrams for the ac (panel a) and dc (panel b) drag current densities in the lowest order in interactions.}
	\label{fig:DCDrag}
\end{figure}

The unperturbed Green functions for the states  near the K-points are determined by the Hamiltonian Eq. (\ref{graphene}). They can be conveniently represented in the form that separates a valley independent pole structure and a valley dependent pseudospin structure 
\begin{equation}
	\hat{G}_{\nu}^{R/A} (\epsilon, {\bf p})=G^{R/A} (\epsilon,{\bf p}) \hat{g}_{\nu}(\epsilon, {\bf p}), 
	\label{NotationsGF}
\end{equation}
where $\nu=K,  K'$ denotes the valley index,  
\begin{equation}
	G^{R/A} (\epsilon,{\bf p}) =\frac{\epsilon}{(\epsilon\pm \frac{i}{2\tau})^2-v_F^2 p^2}
	\label{GF_PoleStructure}
\end{equation}
denote the  pseudospin-independent  parts of the Green functions, and 
\begin{equation}
	\hat{g}_{K/K'}(\epsilon, {\bf p})=\sigma_0+ \frac{v_F}{\epsilon}(p_x\sigma_x\pm p_y\sigma_y).
\end{equation} 
Here the Pauli-matrices $\sigma_i$ act on pseudospin.  In Eq. (\ref{GF_PoleStructure}) we introduced a phenomenological mean-free time $\tau$ that results in the broadening of the poles. 
The Keldysh component of the Green functions is given by 
$\hat{G}_{\nu}^{K} (\epsilon, {\bf p})=\tanh\left(\frac{\epsilon}{2T}\right) \left( \hat{G}_{\nu}^{R} (\epsilon, {\bf p})-  \hat{G}_{\nu}^{A} (\epsilon, {\bf p})\right)$ (here $\nu=K,K'$ is the valley index). The lowest order contributions to the drag current density are given by the diagrams in Fig. (\ref{fig:DCDrag}), where solid lines denote Green functions for the states close to $K$ and $K'$ points. Detailed calculations of AC- and DC- drag current densities  are provided in the Supplemental Material \cite{Supplement}.    

The AC drag current at the frequency $\omega$ is given by the diagrams of the first order in the interaction strength in Fig. \ref{fig:DCDrag}a).  Here the two diagrams correspond to the current induced in the $K$ and $K'$ valleys by inter-valley scattering.  However, we believe that the AC-drag will be strongly suppressed in the realistic experiment because of defects in the ice lattice structure. 
Those defects would lead to phase slips in the driving potential and eventually to a randomization of the phase, which in turn will  suppress the AC drag current. 

In contrast, the DC current is largely insensitive to small random changes in the phase of the driving potential. 
The total DC current density is calculated as a sum of contributions from the two valleys, ${\bf j}^{\mathrm{DC}}= 	\left\langle {\bf j}_{KK}\right\rangle + 	\left\langle {\bf j}_{K'K'}\right\rangle $, each represented by a diagram in Fig. \ref{fig:DCDrag}b).  These diagrams depict the  interaction-induced rectification, where the AC displacement current of the moving layer is converted into a DC drag current in graphene. 

They correspond to the following mathematical expressions 
\begin{eqnarray}
	\nonumber && 
	\left\langle {\bf j}_{KK}\right\rangle =	-i \int_{\epsilon, {\bf p} }\left\{G^R(\epsilon, {\bf p}) G^A(\epsilon, {\bf p}) \left[\tanh\left(\frac{\epsilon+\omega-\mu}{2T}\right) - \tanh\left(\frac{\epsilon-\mu}{2T}\right) \right]\left[G^R(\epsilon+\omega, {\bf p}+{\bf q})- G^A(\epsilon+\omega, {\bf p}+{\bf q}) \right]  + \right. \\ 
	\nonumber && \left. 
	\tanh\left(\frac{\epsilon-\mu}{2T}\right) \left[\left(G^R(\epsilon, {\bf p})\right)^2 G^R(\epsilon+\omega, {\bf p}+{\bf q}) - \left(G^A(\epsilon, {\bf p})\right)^2 G^A(\epsilon+\omega, {\bf p}+{\bf q})  \right]
	\right\} \times \\ 
	&& 
	\mathrm{tr}\left\{ {\bf J}_{KK}({\bf p}, {\bf p}) \hat{g}_K(\epsilon, {\bf p})({\bf u}^*({\bf q}, {\bf p})\cdot{\bm\sigma}) \hat{g}_{K'}(\epsilon+\omega, {\bf p}+{\bf q}) ({\bf u}({\bf q}, {\bf p})\cdot{\bm\sigma}) \hat{g}_K(\epsilon, {\bf p}) 
	\right\}, 
	\label{JKK_GF} \\ 
		\nonumber && 
	\left\langle {\bf j}_{K'K'}\right\rangle =	-i \int_{\epsilon, {\bf p} }\left\{G^R(\epsilon+\omega, {\bf p+q}) G^A(\epsilon+\omega, {\bf p+q}) \left[\tanh\left(\frac{\epsilon-\mu}{2T}\right) - \tanh\left(\frac{\epsilon+\omega-\mu}{2T}\right) \right]\left[G^R(\epsilon, {\bf p})- G^A(\epsilon, {\bf p}) \right] 
	\right\} \times \\ 
	 && 
	\mathrm{tr}\left\{ {\bf J}_{K'K'}({\bf p+q}, {\bf p+q}) \hat{g}_{K'}(\epsilon+\omega, {\bf p+q})({\bf u}({\bf q}, {\bf p+q})\cdot{\bm\sigma}) \hat{g}_{K}(\epsilon, {\bf p}) ({\bf u}^*({\bf q}, {\bf p+q})\cdot{\bm\sigma}) \hat{g}_{K'}(\epsilon+\omega, {\bf p+q}) 
	\right\}. 	\label{JK'K'_GF} 
\end{eqnarray}
Explicit form of the current vertices ${\bf J}_{KK}, {\bf J}_{K'K'}$ is given by 
\begin{equation}
	{\bf J}_{\nu\nu}({\bf p}, {\bf p})= 	\left(\begin{array}{cc} 
		0 &  {\bf j}_{\nu}({\bf p})  \\ 
		{\bf j}^*_{\nu}({\bf p})  & 0 \end{array}\right), 
	\label{Jaa}
\end{equation}
where 
\begin{equation}
	{\bf j}_K({\bf p})=- t \sum_{j=1}^3  {\bm \delta}_j  e^{ i({\bf K}+{\bf p})({\bm \delta}_j-{\bm\delta}_3)  }, \, \, \, \, 	{\bf j}_{K'}({\bf p})=- t \sum_{\langle j \rangle}  {\bm \delta}_j  e^{ i({\bf K'}+{\bf p})({\bm \delta}_j-{\bm\delta}_3)  } 
	\label{jPartial}
\end{equation} 
with ${\bf \delta}_i$ being the nearest neighbor vectors in the graphene lattice, 
\begin{equation}
	{\bm \delta}_1=\frac{a}{2}(1, \sqrt{3}), \, \, \, {\bm \delta}_2=\frac{a}{2}(1, - \sqrt{3}), {\bm \delta}_3=-a(1, 0). 
\end{equation}
Detailed derivation of the current vertices is provided in the Supplemental Material \cite{Supplement}. 
Since the drag requires a finite chemical potential  $\mu$ in graphene, the temperature effects are of minor importance for $T\ll \mu$, which is valid up to room temperature.  
In this case, the calculation of  the diagrams in Fig. \ref{fig:DCDrag}b) can be performed by setting $T=0$. Furthermore, for $\mu>0$  the states in conduction band provide the dominant contribution to the drag current.  This allows simplified calculation of the drag current using the projection on the states of conduction band, which is introduced below. 

\subsection{Projection on the conduction band}

Let $\phi_p$ be the polar angle of the wave vector ${\bf p}=(p_x, p_y)$. Close to the points $K$ and $K'$,  the eigenstates of the conduction band are expressed through the eigenstates of  $A$ and $B$ sublattices as follows \cite{KatsnelsonBook2012} 
\begin{equation}
	|p\rangle_c^K=\frac{1}{\sqrt{2}} \left( e^{-i\phi_{\bf p}/2} |A\rangle + e^{i\phi_{\bf p}/2} |B\rangle\right), \, \, \, \, 
	|p'\rangle_c^{K'}=\frac{1}{\sqrt{2}} \left( e^{i\phi_{\bf p'}/2} |A\rangle + e^{-i\phi_{\bf p'}/2} |B\rangle\right)
\end{equation}
The calculation of the drag current according to Eqs. (\ref{JKK_GF}), (\ref{JK'K'_GF}) involves the matrix element of the current operators 
$
\langle {\bf p},K|	{\bf j}_K({\bf p})	|{\bf p},K\rangle_c
$,  $
\langle {\bf p'},K'|	{\bf j}_{K'}({\bf p'})	|{\bf p'},K'\rangle_c
$. 
The wave vectors ${\bf p}$ and   ${\bf p'}$ are bound by  the kinematic restriction on the wave vectors of the initial and final states due to the energy and quasi-momentum conservation, which we now consider in detail.  
Since the energy transfer by the scattering processes is much smaller than the Fermi energy, we can approximately set the absolute value of the wave vectors for the initial and final states equal to the Fermi wave vector $p_F$.  
It turns out that this condition leaves only two possible choices for the wave vector of the initial state ${\bf p}$ (the wave vector of the final state is then fixed automatically to ${\bf p} +{\bf q}$) at fixed transferred wave vector ${\bf q}$ (see Fig. \ref{fig:Angles}). Denote the polar angle of the wave vector $\bf q$ as $\alpha$ and the angle between vectors ${\bf q}$ and ${\bf p}$ as $\phi$. Then the two possible choices are 
\begin{figure}[t!]
	\centering
	\includegraphics[width=0.5\textwidth]{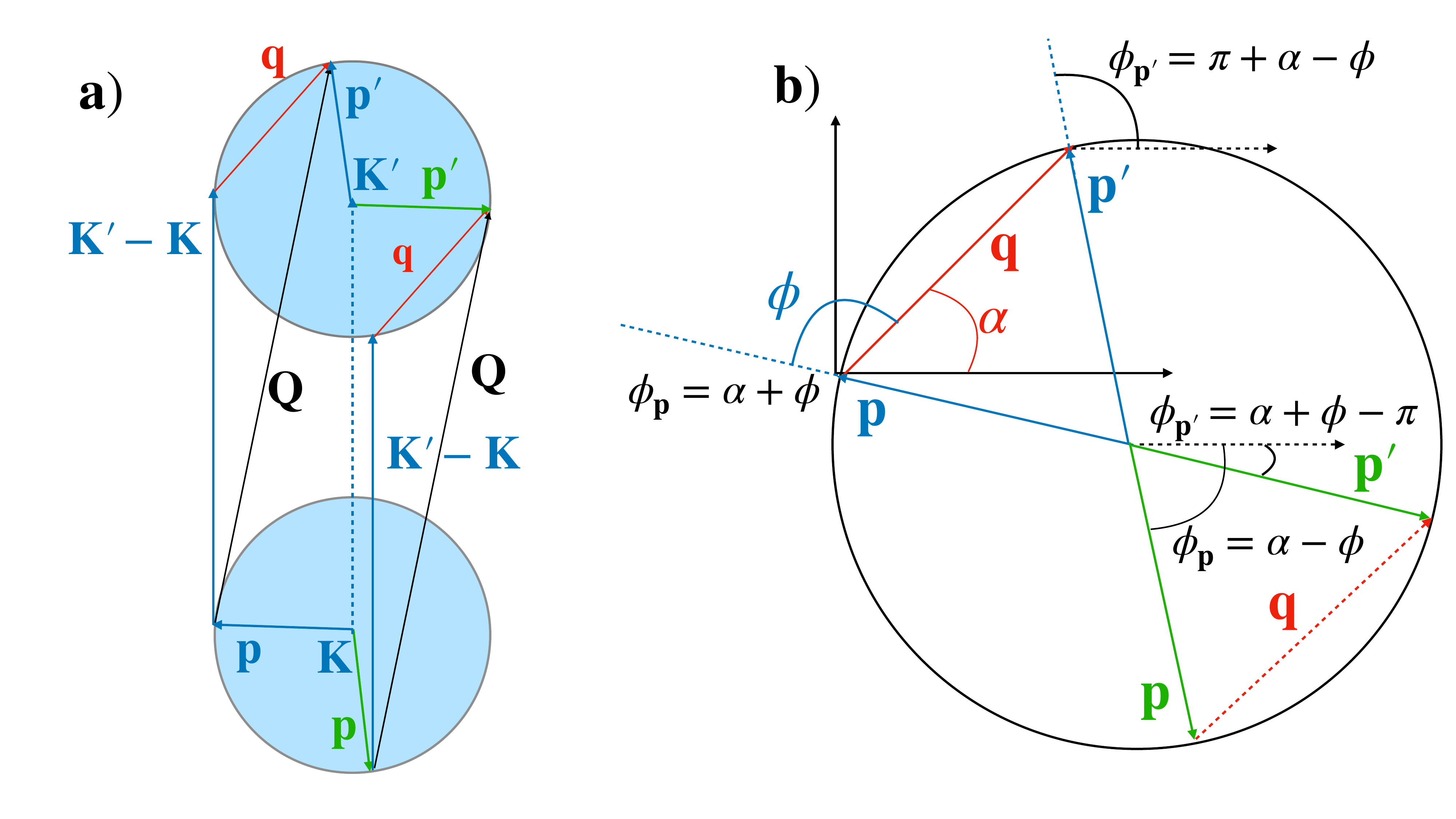}
	\caption{a) Initial and final states by inter-valley scattering for the two possible choices specified by Eqs. (\ref{phiphi'_i}), (\ref{phiphi'_ii}); b) After subtraction of the constant vector ${\bf K'}-{\bf K}$, the Fermi surfaces coincide (the energy difference $\omega$ is neglected).      Definitions of the angles $\alpha$ and $\phi$ and their relation to the directions of the initial (${\bf p}$) and the final (${\bf p'}$) momenta by a single scattering event. At the threshold doping, the vector ${\bf Q}$ is parallel to ${\bf K}-{\bf K'}$, the transferred quasi-momentum $\hbar q=2  p_F$, the angle $\phi=\pm \pi$.  }
	\label{fig:Angles}
\end{figure} 
\begin{eqnarray}
	&& 
	{\bf i)}\, \, \, 	{\phi_{\bf p}}=\alpha+\phi, \, \, \, {\phi_{\bf p'}}=\pi+ \alpha-\phi,  \label{phiphi'_i} \\ 
	&& 
	{\bf ii)}\, \, \, 
	{\phi_{\bf p}}=\alpha-\phi, \, \, \, {\phi_{\bf p'}}= \alpha+\phi-\pi <=> \pi + \alpha+\phi. \label{phiphi'_ii}
\end{eqnarray}
It follows that all results for the case (ii) are obtained from (i) by the change $\phi\rightarrow - \phi$. The change of the sign of the transfer quasi-momentum ${\bf q}\rightarrow -{\bf q}$ corresponds to the change $\alpha\rightarrow \alpha+ \pi$.  

Let us write down explicit expression for the components of current density operators Eqs. (\ref{jPartial}) projected on the conduction band. We consider the geometry as given in the case {\bf (i)} above. 
\begin{eqnarray}
	&& {\bf j}_K({\bf p})=-ta\left\{ \left(\begin{array}{c}
		-1 \\ 0
	\end{array}\right)+ 
	\left(\begin{array}{c}
		1/2 \\ \sqrt{3}/2
	\end{array}\right) e^{i\frac{2\pi}{3}} e^{i\sqrt{3}a p_F\cos(\alpha+\phi-\pi/6)} 
	+  	\left(\begin{array}{c}
		1/2 \\ -\sqrt{3}/2
	\end{array}\right) e^{-i\frac{2\pi}{3}} e^{i\sqrt{3}a p_F\cos(\alpha+\phi+\pi/6)}  \right\} \label{jK_p} \\ 
	\nonumber 	&& {\bf j}_{K'}({\bf p}+{\bf q})=  -ta\left\{ \left(\begin{array}{c}
		-1 \\ 0
	\end{array}\right)+ 
	\left(\begin{array}{c}
		1/2 \\ \sqrt{3}/2
	\end{array}\right) e^{-i\frac{2\pi}{3}} e^{-i\sqrt{3}a p_F\cos(\alpha-\phi-\pi/6)} 
	+  	\left(\begin{array}{c}
		1/2 \\ -\sqrt{3}/2
	\end{array}\right) e^{i\frac{2\pi}{3}} e^{-i\sqrt{3}a p_F\cos(\alpha-\phi+\pi/6)}  \right\} \\ 
	\label{jK'_p'} 
\end{eqnarray}
The matrix elements of the current density operators between the states in conduction band  are given by 
\begin{eqnarray}
	\nonumber 	&& 
	\langle {\bf p}, K|{\bf J}_{KK}({\bf p}, {\bf p})|{\bf p}, K\rangle_c= 
	\frac{1}{2} \left(e^{i\phi/2}, e^{-i\phi/2} \right) \left(\begin{array}{cc} 
		0 & {\bf j}_K({\bf p}) \\ 
		{\bf j}_{K}^*({\bf p}) & 0  
	\end{array} \right) 
	\left(\begin{array}{c}
		e^{-i\phi/2} \\  e^{i\phi/2} 
	\end{array}\right) = \\ 
	&& 
	\frac{1}{2}\left\{ {\bf j}_K({\bf p}) e^{i\phi}  + {\bf j}_{K}^*({\bf p}) e^{-i\phi} \right\}, 
	\label{JKK_projected}
\end{eqnarray}
where $\phi=\phi_{\bf p}$. 
Analogously, for the valley $K'$ we obtain 
\begin{eqnarray}
	\nonumber 	&& 
	\langle {\bf p'}, K'|{\bf J}_{K'K'}({\bf p'}, {\bf p'})|{\bf p'}, K'\rangle_c= 
	\frac{1}{2} \left(e^{-i\phi'/2}, e^{i\phi'/2} \right) \left(\begin{array}{cc} 
		0 & {\bf j}_{K'}({\bf p'}) \\ 
		{\bf j}_{K'}^*({\bf p'}) & 0  
	\end{array} \right) 
	\left(\begin{array}{c}
		e^{i\phi'/2} \\  e^{-i\phi'/2} 
	\end{array}\right) = \\ 
	&& 
	\frac{1}{2}\left\{ {\bf j}_{K'}({\bf p'}) e^{-i\phi'}  + {\bf j}_{K'}^*({\bf p'}) e^{i\phi'} \right\}, 
	\label{JK'K'_projected}
\end{eqnarray}
where $\phi'=\phi_{{\bf p}+{\bf q}}$ as given by Eqs. (\ref{phiphi'_i}), (\ref{phiphi'_ii}).  
Matrix elements of the projected interaction vertices are given by 
\begin{eqnarray}
	&& \langle {\bf p+q}, K'| u_0\sigma_0| {\bf p},K\rangle_c= -\frac{1}{2} e^{-\frac{i}{2}|q| a \cos\alpha} \cos\left(\frac{|q| a}{2} \cos\alpha\right) \sin\alpha, \label{u0}\\ 
	&& \langle {\bf p+q}, K'| u_3\sigma_z| {\bf p},K\rangle_c= \frac{1}{2} e^{-\frac{i}{2}|q| a \cos\alpha} \sin\left(\frac{|q| a}{2} \cos\alpha\right) \cos\alpha, \label{u3} \\ 
\nonumber 	&& \langle {\bf  p+q}, \! K'| u_1\sigma_x| {\bf p},\! K\rangle_c\!=\! \frac{U_1}{2 U_0} e^{-\frac{i}{2}|q| a \cos\alpha} \sin\phi \left\{1\!-\! \cos\left[ p_F a\sin\phi \cos\left(\frac{\pi}{3}\! -\! \alpha\right)\right]\! - \! \cos\left[ p_F a\sin\phi \cos\left(\frac{2\pi}{3}\! - \! \alpha\right)\right]  \right\}, \\ 
	&& \label{u1} \\ 
\nonumber 	&& \langle {\bf p+q}, K'| u_2\sigma_y| {\bf p},K\rangle_c= -\frac{U_1}{2 U_0} e^{-\frac{i}{2}|q| a \cos\alpha} \cos\phi \left\{\sin\left[ p_F a\sin\phi \cos\left(\frac{\pi}{3}\!-\!\alpha\right)\right]\! + \!  \sin\left[ p_F a\sin\phi \cos\left(\frac{2\pi}{3}\! -\! \alpha\right)\right] \right\}. \\ 
&&  \label{u2}
\end{eqnarray}

Upon projection on the conduction band, the leading contribution to the DC drag current density is given by 
\begin{eqnarray}
	\nonumber && 
 {\bf j}^{\mathrm{DC}}=	\left\langle {\bf j}_{KK}\right\rangle + 	\left\langle {\bf j}_{K'K'}\right\rangle  =- \frac{\pi^2 \tau}{2} U_0^2 \int_{{\bf p} }  \left( \langle {\bf p}, K|{\bf J}_{KK}({\bf p}, {\bf p})|{\bf p}, K\rangle_c -  \langle {\bf p+q}, K'|{\bf J}_{K'K'}({\bf p+q}, {\bf p+q})| {\bf p+q}, K'\rangle_c \right) \times\\ 
\nonumber 
 && 
   \langle {\bf p}, K| ({\bf u}^*\cdot{\bm{\sigma}})|{\bf p}+{\bf q}, K'\rangle_c \langle {\bf p}+{\bf q}, K'|  ({\bf u}\cdot{\bm{\sigma}}) | {\bf p}, K\rangle_c 	\left[\tanh\left(\frac{v_F|p|+\omega-\mu}{2T}\right) - \tanh\left(\frac{v_F|p|-\mu}{2T}\right) \right]
	\delta(v_F|p+q|-v_F |p| -\omega)\approx \\ 
\nonumber 	&& 
	 \frac{\pi^2 \tau \omega}{v_F} U_0^2  \int_{{\bf p} }  \left( \langle {\bf p}, K|{\bf J}_{KK}({\bf p}, {\bf p})|{\bf p}, K\rangle_c -  \langle {\bf p+q}, K'|{\bf J}_{K'K'}({\bf p+q}, {\bf p+q})| {\bf p+q}, K'\rangle_c \right) \times\\ 
	 && 
	 \langle {\bf p}, K| ({\bf u}^*\cdot{\bm{\sigma}})|{\bf p}+{\bf q}, K'\rangle_c \langle {\bf p}+{\bf q}, K'|  ({\bf u}\cdot{\bm{\sigma}}) | {\bf p}, K\rangle_c  \delta(|p|-p_F) 	\delta(v_F|p+q|-v_F |p| -\omega), 
	\label{JDCprojected}
\end{eqnarray} 
where in last equation we used the approximation 
\begin{equation}
	\tanh\left(\frac{v_F|p|+\omega-\mu}{2T}\right) - \tanh\left(\frac{v_F|p|-\mu}{2T}\right) = 2\left[f\left(\frac{v_F|p|-\mu}{T}\right) - f\left(\frac{v_F|p|+\omega-\mu}{T}\right) \right]\approx 2 \frac{\omega}{v_F } \frac{\partial f}{\partial|p|}\approx 2 \frac{\omega}{v_F }\delta(|p|-p_F), 
\end{equation}
valid for $T\ll \omega\ll \mu$. 

 It is convenient to perform the two-dimensional integration over momenta using polar coordinates, $p=(|{\bf p}|, \phi)$ where $\phi$ is defined as the angle between  vectors ${\bf p}$ and ${\bf q}$. Furthermore, we satisfy the $\delta$-functions by performing the angular integration. The angular integrals are calculated according to the formula 
\begin{equation}
	\int d\phi \delta(f(\phi)) F(\phi)=\frac{F(\phi_0)}{|f'(\phi_0)|}, 
\end{equation}
where $f'(\phi_0)=\frac{df(\phi)}{d\phi}\vert_{\phi=\phi_0}$, and $\phi_0$ satisfies the relation $f(\phi_0)=0$.
Here we show explicitly the computation of the integral with $\delta(v_F|p+q|-v_F|p|-\omega)$ in Eq. (\ref{JDCprojected}). The angular dependence in the $\delta$-function is contained in the term 
\begin{equation}
	|p+q|=\sqrt{p^2+q^2+2pq\cos\phi}, 
\end{equation}
therefore 
\begin{equation}
	f(\phi)=\omega+v_F p - v_F\sqrt{p^2+q^2+2pq\cos\phi}, \, \, \, \, 
	f'(\phi)=\frac{v_F pq\sin\phi}{\sqrt{p^2+q^2+2pq\cos\phi}}=\frac{v_F pq\sin\phi}{|p+q|}= 
	\frac{v_F^2 pq\sin\phi}{\omega+v_F p}, 
\end{equation}
where in the last equation we used the condition imposed by the $\delta$-function. 	Furthermore, in the physically relevant regime $v_F p\gg \omega$, we can simplify  
\begin{equation}
	f'(\phi)\approx v_F q\sin\phi. 
\end{equation}
At low temperature, $T\ll \omega$, the $\tanh$ can be replaced by a step function. Then, the integration over the absolute value of momentum $|p|$ is performed by replacing $|p|\approx p_F$, whereas the difference of the step-functions  determines the integration range $\omega/v_F$. Thus, the low-temperature approximation to the integration can be summarized by the expression 
\begin{equation}
	\int_{\bf  p}  \left[\tanh\left(\frac{\omega+ v_F |p|-\mu}{2T}\right)- \tanh\left(\frac{v_F |p| -\mu}{2T}\right)  \right] \delta(v_F|{\bf p+q}|-\omega-v_F|p|) ... \approx 
	\frac{\omega p_F}{v_F^2 |q|}  \int d|p| \int \frac{d\phi}{\sin\phi} \delta(|p|-p_F)\delta(\phi-\phi_0) ... , 
\end{equation}
where $\phi_0$  is the angle between $\bf p$ and $\bf q$ that solves the equation  
$|{\bf p+q|}=|p|=p_F$, which is  explicitly  given by 
\begin{equation}
	\phi_0=\arccos \left(-\frac{q}{2p_F}\right). 
	\label{phi_0}
\end{equation} 

\subsection{DC-Current: results and discussion} 
Performing the integration over momentum, we finally obtain 
the following expression for the  DC-drag current density above the threshold chemical potential 
\begin{eqnarray}
	\nonumber && 
	{\bf j}^{\mathrm{DC}}= 
	\frac{\pi^2 U_0^2 \tau \omega p_F}{v_F^2 |q||\sin\phi|}   
	\left[
	\langle {\bf p}, K|{\bf J}_{KK}({\bf p}, {\bf p})| {\bf p}, K\rangle_c - 	\langle {\bf p+q}, K'|{\bf J}_{K'K'}({\bf p+q}, {\bf p+q})| {\bf p+q}, K'\rangle_c \right] \times \\ 
\nonumber 	&& 
	\langle {\bf p}, K| ({\bf u}^*\cdot{\bm{\sigma}})|{\bf p}+{\bf q}, K'\rangle_c \langle {\bf p}+{\bf q}, K'|  ({\bf u}\cdot{\bm{\sigma}}) | {\bf p}, K\rangle_c  =    \frac{\pi^2 e}{\hbar^3} \frac{U_0^2 p_F\omega \tau}{2 v_F|q|} \frac{\mathcal{\bf F}^{\mathrm{DC}}(\alpha, \phi) }{|\sin\phi|} 	=	
	 \frac{\pi^4}{e\hbar}\frac{v_0}{a_0} \frac{U_0^2}{\rho_g v_F|q|} \frac{\mathcal{\bf F}^{\mathrm{DC}}(\alpha, \phi) }{|\sin\phi|}.    \\ 
	 \label{j_dc}
\end{eqnarray} 
In the last equation, we expressed  the mean free time $\tau$ through the resistivity of graphene $\rho_g=\frac{\pi \hbar}{e^2\mu \tau}$, where $\mu=p_Fv_F$ denotes the chemical potential. Additionally, we used the relation between the frequency and the velocity of the crystalline layer  $\omega=2\pi v_0/a_0$,  assuming the velocity to be parallel to the  vector ${\bf Q}$ of the moving polar crystal.  

\begin{figure}[t!]
	\centering
	\includegraphics[width=\textwidth]{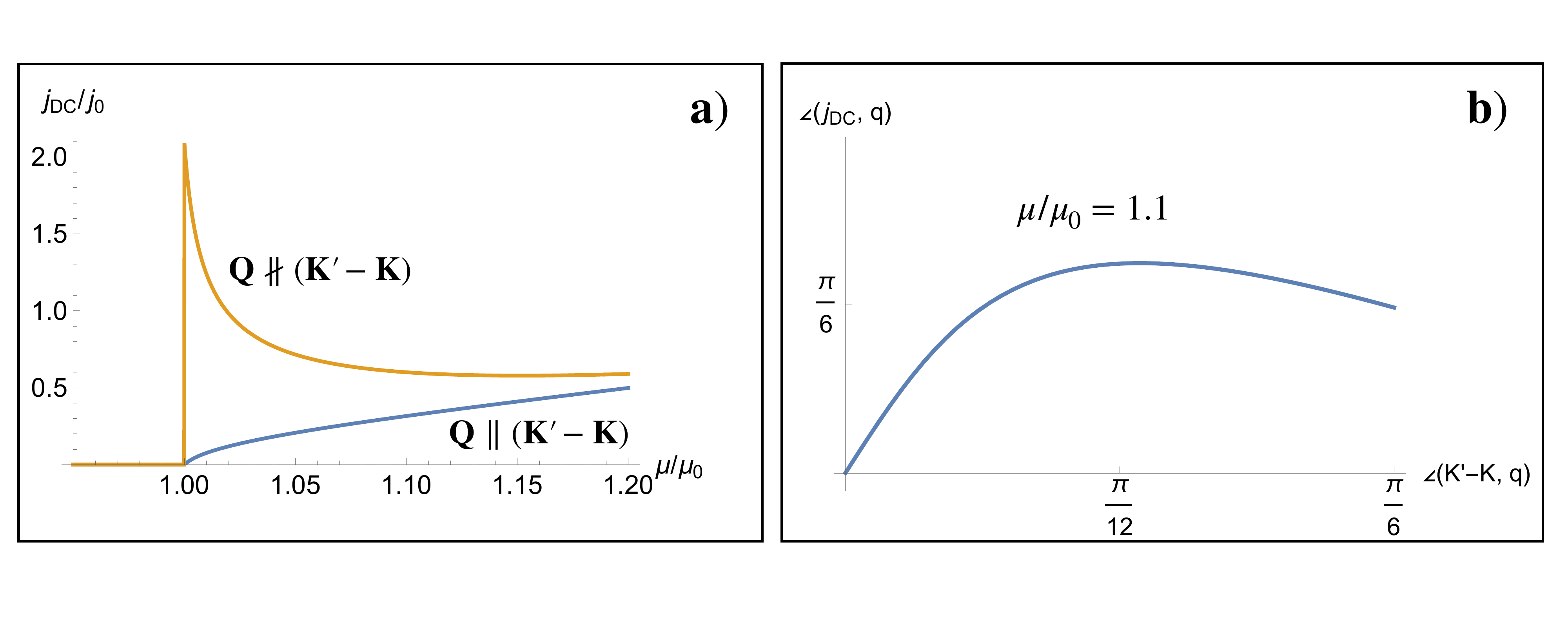}
	\vskip -.3cm
	\caption{a) DC drag current as  function of chemical potential for two different directions of the transferred momentum ${\bf Q}$.  Except for the momentum ${\bf Q}$ parallel to the vector ${\bf K'}-{\bf K}$ connecting the two $K$-points, there is a singularity in the drag current at the threshold chemical potential (the curve for the angle $\angle ({\bf Q}, {\bf K'}-{\bf K'})=\pi/10$ is shown). The current starts growing continuously  at the threshold for ${\bf Q}$ parallel to ${\bf K'}-{\bf K}$. The threshold chemical potential $\mu_0$ corresponds to the  transferred quasi-momentum $q=2 p_F$. The normalization $j_0= \frac{\pi^4}{e\hbar}\frac{v_0}{a_0} \frac{U_0^2}{\rho_g v_F|q|}$ is independent of the chemical potential; b)  The angle between the direction of the transferred quasi-momentum ${\bf q}={\bf Q}-({\bf K'}- {\bf  K})$ and the drag current ${\bf j}_{\mathrm{DC}}$  as  function of the angle between ${\bf q}$ and ${\bf K'}- {\bf  K}$.  In the absence of the nonlinear Hall effect, the drag current would be parallel to ${\bf q}$. }
	\label{fig:DCCurrent}
\end{figure}

The dependencies of the drag current on chemical potential for two different directions of the transferred quasi-momentum ${\bf q}$ are shown in Fig. \ref{fig:DCCurrent}a).   There is a sharp threshold chemical potential for the drag effect. The drag current given by Eq. (\ref{j_dc}) formally diverges at the threshold $p_F=\frac{1}{2}(q-\omega/v_F)$, corresponding to the angle $\phi=\pi$. This divergence is removed by accounting for scattering processes that change the momentum of quasiparticles.  
Considering the finite lifetime $\tau$  leads to the replacement $\sin\phi\rightarrow \frac{1}{\sqrt{v_F q\tau}}$ in Eq. (\ref{j_dc}) in the region $v_F|2p_F-q| <1/\tau$ close to the threshold.  Finite temperature leads to a smearing of the threshold  $\mu= \left(\frac{v_F |q|}{\omega}-1\right)/2$, but it does affect the qualitative dependence of the current on chemical potential.  Above the threshold, the direction of the current is determined by the vector ${\bf F}^{\mathrm{DC}}(\alpha, \phi)$, which is given by the product of matrix elements of the current density operator and interaction vertices 
\begin{equation}
	{\bf F}^{\mathrm{DC}}(\alpha, \phi)=\Re[{\bf j}_K({\bf p})e^{i\phi_{\bf p}} -{\bf j}_{K'}({\bf p})e^{i\phi_{\bf p'}} ] \left|\langle {\bf p+q},K'|({\bf u}\cdot\bm{\sigma})|{\bf p},K\rangle\right|^2,
	\label{FDC}
\end{equation}
where $\phi_{\bf p}= \alpha\pm \phi$ denotes the direction of the vector ${\bf p}$, and $\phi_{\bf p'}= \alpha\mp \phi$ denotes the direction of the vector ${\bf p'}$,  as shown in Fig. \ref{fig:Angles}.  

Eq. (\ref{j_dc}) allows an interpretation of the dc-drag in terms of the pseudospin dipole moment induced by the inter-valley scattering. Namely, the current density operator in graphene is intimately related to the pseudospin of the current-carrying state. For instance, for the states close to $K$ and $K'$ points the current density operator is simply proportional to the pseudospin, ${\bf j}_K=v_F {\bm \sigma}$, ${\bf j}_{K'}=v_F {\bm \sigma^*}$. The drag potential of the moving crystalline layer excites inter-valley particle-hole pairs at the states $|{\bf p},K\rangle$, $|{\bf p+q},K'\rangle$, which in turn leads to a deviation of the average pseudospin at these states from its equilibrium value. Because of the opposite sign of nonequilibrium charges created in the different values, the drag current is proportional to the {\em difference} of the nonequilibrium pseudospins, $\langle {\bf p}, K|{\bf J}_{KK}({\bf p}, {\bf p})| {\bf p}, K\rangle_c - 	\langle {\bf p+q}, K'|{\bf J}_{K'K'}({\bf p+q}, {\bf p+q})| {\bf p+q}, K'\rangle_c$. The direction of the nonequilibrium pseudospin does not coinside in general with the direction of the transfered quasimomentum $\bf q$. Therefore, contrary to naive expectation, the  direction of the drag current deviates from the direction of transfered quasimomentum, as shown in Fig. \ref{fig:DCCurrent}b). This deviation can be interpreted as  a signature of the nonlinear Hall effect, since the DC drag current is proportional to the second order of the driving potential, and hence to the second order of the driving electric field.   

\section{Drag-induced viscosity} 
The drag effect considered here represents the conversion of the kinetic energy of the moving polar crystalline layer into the energy of a DC electric current in graphene. 
This energy transfer from the moving crystalline layer to the  electrons in graphene results in the back-action force on the layer, which can be formulated in terms of a renormalized shear viscosity coefficient.   If one adopts the Poiseuille equation for a crystalline layer moving with a velocity $v_0$ between the two graphene layers, the following relation is obtained
$
	\Delta p h L_{\perp} =8 \eta v_0 L_{\perp}L/h 
$, 
where $\Delta p$ denotes the pressure difference at the ends of the interlayer channel of the length $L$, $h$ denotes the distance between the graphene layers forming the channel (see Fig. \ref{fig:Transitions}a), $v_0$ is the velocity of the crystalline layer, and $\nu_0$ denotes the shear viscosity in the absence of the drag current. To determine the change in shear viscosity due to the drag, consider the energy dissipation during the motion of the crystalline layer. 
In the absence of the drag current, the energy dissipation rate is given by the relation 
$
	\dot{Q}_0=	\Delta p h L_{\perp} v_0= 8 \eta_0 v_0^2 L_{\perp}L/h. 
$ 
The drag current results in the additional dissipation, given by 
$
	\dot{Q}_{\mathrm{drag}}= 2 j_d^2 L_{\perp}^2\rho_g L= f_d^2 v_0^2 L_{\perp}^2\rho_g L,
$
where $j_d=v_0 f_d$ is the absolute value of the drag current density in graphene, $\rho_g$ denotes the resistivity of the graphene layer, and the factor 2 accounts for the two layers forming the nano-channel. In the last equation, we took into account the proportionality of the drag current to the velocity of the moving layer, as given by Eq. (\ref{j_dc}). The coefficient $f_d$ can be read from Eq. (\ref{j_dc}),  $f_d= \frac{\pi^4}{e\hbar}\frac{U_0^2}{a_0 \rho_g v_F|q|} \frac{|\mathcal{\bf F}^{\mathrm{DC}}(\alpha, \phi)| }{|\sin\phi|}$.  Therefore, the total energy dissipation in presence of the drag reads 
\begin{equation}
	\dot{Q}_0+	\dot{Q}_{\mathrm{drag}}=\eta_0 v_0^2 \frac{8L_{\perp}L}{h}+f_d^2 v_0^2 L_{\perp}^2\rho_g L\equiv \eta v_0^2 \frac{8L_{\perp}L}{h}, 
	\label{Dissipation_drag}
\end{equation}
where the last equation defines the renormalized shear friction coefficient $\eta$. Eq. (\ref{Dissipation_drag}) allows one to determine the shear friction coefficient due to the drag effect, $\eta = \eta_0+ \eta_{d}$, where 
$
	\eta_{d}=f_d^2  \rho_g  L_{\perp} h/4
$. 
The change in dynamic viscosity provides an alternative way to detect the drag phenomenon. Specifically, crossing the drag threshold by altering the chemical potential of graphene would lead to a sharp decrease in the velocity of the crystalline layer from $v$ to $v_0$, as described by the relation 
$
	\frac{v_0}{v}=\frac{\eta_0}{\eta_0+\eta_d}. 
$

\section{Conclusion} 
In summary, we predict a Coulomb drag effect that should be observable in experiments involving sliding polar crystalline layers through nano-constrictions between graphene planes. This drag mechanism hinges on the comparable lattice constants of graphene and the moving crystalline layer. In the electrotechnical framework of Coulomb drag, the moving polar crystal serves as an active layer that generates a driving electric field with an extremely short wavelength. This wave length is comparable with microscopic lengths of the passive layer (specifically graphene), which allows probe of excitations created by inter-valley and/or umklapp scattering events. This feature is unique and usually unreachable for the drag by phonons or surface acoustic waves. 
We believe that the Coulomb drag by moving polar crystalline layer can be used as a probe of excitations by umklapp and inter-valley scattering in layered materials  beyond graphene. 

The predicted effect can be experimentally detected through the induction of drag current or drag voltage in graphene planes, as well as by observing an increase in viscosity due to the motion of the crystalline layer, resulting in a reduction of its velocity. Additionally, the drag exhibits signatures of the nonlinear Hall effect—specifically, a component of the drag current perpendicular to the transferred quasi-momentum.  As it was shown in Refs.  \cite{deyo2009,Sodemann2015}, the nonlinear Hall effect serves as a probe of the Berry curvature dipole of the medium. This observation rises a curiosity for the relation of Hall current  to the Berry curvature dipole in graphene.    Achieving the drag effect requires significant doping of graphene to create sufficiently large Fermi surfaces around each $K$-point. Our estimations indicate that modern experimental doping levels are sufficient for observing this phenomenon.

\section*{Acknowledgements}
The author acknowledges fruitful discussions with M. Trushin, A. Kamenev, A. Levchenko, V. Cheianov, T. Wehling, and M. Khodas. 

\bibliography{WaterDragGraphene.bib}

\begin{thebibliography}{30}%
\makeatletter
\providecommand \@ifxundefined [1]{%
 \@ifx{#1\undefined}
}%
\providecommand \@ifnum [1]{%
 \ifnum #1\expandafter \@firstoftwo
 \else \expandafter \@secondoftwo
 \fi
}%
\providecommand \@ifx [1]{%
 \ifx #1\expandafter \@firstoftwo
 \else \expandafter \@secondoftwo
 \fi
}%
\providecommand \natexlab [1]{#1}%
\providecommand \enquote  [1]{``#1''}%
\providecommand \bibnamefont  [1]{#1}%
\providecommand \bibfnamefont [1]{#1}%
\providecommand \citenamefont [1]{#1}%
\providecommand \href@noop [0]{\@secondoftwo}%
\providecommand \href [0]{\begingroup \@sanitize@url \@href}%
\providecommand \@href[1]{\@@startlink{#1}\@@href}%
\providecommand \@@href[1]{\endgroup#1\@@endlink}%
\providecommand \@sanitize@url [0]{\catcode `\\12\catcode `\$12\catcode
  `\&12\catcode `\#12\catcode `\^12\catcode `\_12\catcode `\%12\relax}%
\providecommand \@@startlink[1]{}%
\providecommand \@@endlink[0]{}%
\providecommand \url  [0]{\begingroup\@sanitize@url \@url }%
\providecommand \@url [1]{\endgroup\@href {#1}{\urlprefix }}%
\providecommand \urlprefix  [0]{URL }%
\providecommand \Eprint [0]{\href }%
\providecommand \doibase [0]{https://doi.org/}%
\providecommand \selectlanguage [0]{\@gobble}%
\providecommand \bibinfo  [0]{\@secondoftwo}%
\providecommand \bibfield  [0]{\@secondoftwo}%
\providecommand \translation [1]{[#1]}%
\providecommand \BibitemOpen [0]{}%
\providecommand \bibitemStop [0]{}%
\providecommand \bibitemNoStop [0]{.\EOS\space}%
\providecommand \EOS [0]{\spacefactor3000\relax}%
\providecommand \BibitemShut  [1]{\csname bibitem#1\endcsname}%
\let\auto@bib@innerbib\@empty
\bibitem [{\citenamefont {Ali}\ \emph {et~al.}(2024)\citenamefont {Ali},
  \citenamefont {{Zenab Hasan}}, \citenamefont {Garcia}, \citenamefont
  {Danquah},\ and\ \citenamefont {Imanova}}]{GrapheneMembranes_2024}%
  \BibitemOpen
  \bibfield  {author} {\bibinfo {author} {\bibfnamefont {I.}~\bibnamefont
  {Ali}}, \bibinfo {author} {\bibfnamefont {S.}~\bibnamefont {{Zenab Hasan}}},
  \bibinfo {author} {\bibfnamefont {H.}~\bibnamefont {Garcia}}, \bibinfo
  {author} {\bibfnamefont {M.~K.}\ \bibnamefont {Danquah}},\ and\ \bibinfo
  {author} {\bibfnamefont {G.}~\bibnamefont {Imanova}},\ }\bibfield  {title}
  {\bibinfo {title} {Recent advances in graphene-based nano-membranes for
  desalination},\ }\href
  {https://doi.org/https://doi.org/10.1016/j.cej.2024.149108} {\bibfield
  {journal} {\bibinfo  {journal} {Chemical Engineering Journal}\ }\textbf
  {\bibinfo {volume} {483}},\ \bibinfo {pages} {149108} (\bibinfo {year}
  {2024})}\BibitemShut {NoStop}%
\bibitem [{\citenamefont {Günay}\ \emph {et~al.}(2023)\citenamefont {Günay},
  \citenamefont {Kemerli}, \citenamefont {Karaman}, \citenamefont {Karaman},
  \citenamefont {Güngör},\ and\ \citenamefont
  {Karimi-Maleh}}]{GUNAY2023114785}%
  \BibitemOpen
  \bibfield  {author} {\bibinfo {author} {\bibfnamefont {M.~G.}\ \bibnamefont
  {Günay}}, \bibinfo {author} {\bibfnamefont {U.}~\bibnamefont {Kemerli}},
  \bibinfo {author} {\bibfnamefont {C.}~\bibnamefont {Karaman}}, \bibinfo
  {author} {\bibfnamefont {O.}~\bibnamefont {Karaman}}, \bibinfo {author}
  {\bibfnamefont {A.}~\bibnamefont {Güngör}},\ and\ \bibinfo {author}
  {\bibfnamefont {H.}~\bibnamefont {Karimi-Maleh}},\ }\bibfield  {title}
  {\bibinfo {title} {Review of functionalized nano porous membranes for
  desalination and water purification: Md simulations perspective},\ }\href
  {https://doi.org/https://doi.org/10.1016/j.envres.2022.114785} {\bibfield
  {journal} {\bibinfo  {journal} {Environmental Research}\ }\textbf {\bibinfo
  {volume} {217}},\ \bibinfo {pages} {114785} (\bibinfo {year}
  {2023})}\BibitemShut {NoStop}%
\bibitem [{\citenamefont {Zhu}\ \emph {et~al.}(2019)\citenamefont {Zhu},
  \citenamefont {Wang}, \citenamefont {Fan}, \citenamefont {Xu},\ and\
  \citenamefont {Yang}}]{NanoConfinedChannels2019}%
  \BibitemOpen
  \bibfield  {author} {\bibinfo {author} {\bibfnamefont {H.}~\bibnamefont
  {Zhu}}, \bibinfo {author} {\bibfnamefont {Y.}~\bibnamefont {Wang}}, \bibinfo
  {author} {\bibfnamefont {Y.}~\bibnamefont {Fan}}, \bibinfo {author}
  {\bibfnamefont {J.}~\bibnamefont {Xu}},\ and\ \bibinfo {author}
  {\bibfnamefont {C.}~\bibnamefont {Yang}},\ }\bibfield  {title} {\bibinfo
  {title} {Structure and transport properties of water and hydrated ions in
  nano-confined channels},\ }\href
  {https://doi.org/https://doi.org/10.1002/adts.201900016} {\bibfield
  {journal} {\bibinfo  {journal} {Advanced Theory and Simulations}\ }\textbf
  {\bibinfo {volume} {2}},\ \bibinfo {pages} {1900016} (\bibinfo {year}
  {2019})},\ \Eprint
  {https://arxiv.org/abs/https://onlinelibrary.wiley.com/doi/pdf/10.1002/adts.201900016}
  {https://onlinelibrary.wiley.com/doi/pdf/10.1002/adts.201900016} \BibitemShut
  {NoStop}%
\bibitem [{\citenamefont {Algara-Siller}\ \emph {et~al.}(2015)\citenamefont
  {Algara-Siller}, \citenamefont {Lehtinen}, \citenamefont {Wang},
  \citenamefont {Nair}, \citenamefont {Kaiser}, \citenamefont {Wu},
  \citenamefont {Geim},\ and\ \citenamefont {Grigorieva}}]{SquareIce_Geim2015}%
  \BibitemOpen
  \bibfield  {author} {\bibinfo {author} {\bibfnamefont {G.}~\bibnamefont
  {Algara-Siller}}, \bibinfo {author} {\bibfnamefont {O.}~\bibnamefont
  {Lehtinen}}, \bibinfo {author} {\bibfnamefont {F.~C.}\ \bibnamefont {Wang}},
  \bibinfo {author} {\bibfnamefont {R.~R.}\ \bibnamefont {Nair}}, \bibinfo
  {author} {\bibfnamefont {U.}~\bibnamefont {Kaiser}}, \bibinfo {author}
  {\bibfnamefont {H.~A.}\ \bibnamefont {Wu}}, \bibinfo {author} {\bibfnamefont
  {A.~K.}\ \bibnamefont {Geim}},\ and\ \bibinfo {author} {\bibfnamefont
  {I.~V.}\ \bibnamefont {Grigorieva}},\ }\bibfield  {title} {\bibinfo {title}
  {Square ice in graphene nanocapillaries},\ }\href
  {https://doi.org/10.1038/nature14295} {\bibfield  {journal} {\bibinfo
  {journal} {Nature}\ }\textbf {\bibinfo {volume} {519}},\ \bibinfo {pages}
  {443} (\bibinfo {year} {2015})}\BibitemShut {NoStop}%
\bibitem [{\citenamefont {Negi}\ \emph {et~al.}(2022)\citenamefont {Negi},
  \citenamefont {Carvalho}, \citenamefont {Trushin},\ and\ \citenamefont
  {Neto}}]{Trushin2022}%
  \BibitemOpen
  \bibfield  {author} {\bibinfo {author} {\bibfnamefont {S.}~\bibnamefont
  {Negi}}, \bibinfo {author} {\bibfnamefont {A.}~\bibnamefont {Carvalho}},
  \bibinfo {author} {\bibfnamefont {M.}~\bibnamefont {Trushin}},\ and\ \bibinfo
  {author} {\bibfnamefont {A.~H.~C.}\ \bibnamefont {Neto}},\ }\bibfield
  {title} {\bibinfo {title} {Edge-driven phase transitions in 2d ice},\ }\href
  {https://doi.org/10.1021/acs.jpcc.2c04492} {\bibfield  {journal} {\bibinfo
  {journal} {The Journal of Physical Chemistry C}\ }\textbf {\bibinfo {volume}
  {126}},\ \bibinfo {pages} {16006} (\bibinfo {year} {2022})}\BibitemShut
  {NoStop}%
\bibitem [{\citenamefont {Gao}\ \emph {et~al.}(2018)\citenamefont {Gao},
  \citenamefont {Giovambattista},\ and\ \citenamefont {Sahin}}]{Sahin2018}%
  \BibitemOpen
  \bibfield  {author} {\bibinfo {author} {\bibfnamefont {Z.}~\bibnamefont
  {Gao}}, \bibinfo {author} {\bibfnamefont {N.}~\bibnamefont
  {Giovambattista}},\ and\ \bibinfo {author} {\bibfnamefont {O.}~\bibnamefont
  {Sahin}},\ }\bibfield  {title} {\bibinfo {title} {Phase diagram of water
  confined by graphene},\ }\href {https://doi.org/10.1038/s41598-018-24358-3}
  {\bibfield  {journal} {\bibinfo  {journal} {Scientific Reports}\ }\textbf
  {\bibinfo {volume} {8}},\ \bibinfo {pages} {6228} (\bibinfo {year}
  {2018})}\BibitemShut {NoStop}%
\bibitem [{\citenamefont {Li}\ and\ \citenamefont
  {Schmidt}(2019)}]{li2019twodimensional}%
  \BibitemOpen
  \bibfield  {author} {\bibinfo {author} {\bibfnamefont {S.}~\bibnamefont
  {Li}}\ and\ \bibinfo {author} {\bibfnamefont {B.}~\bibnamefont {Schmidt}},\
  }\bibfield  {title} {\bibinfo {title} {Replica exchange md simulations of
  two-dimensional water in graphene nanocapillaries: rhombic versus square
  structures{,} proton ordering{,} and phase transitions},\ }\href
  {https://doi.org/10.1039/C9CP00849G} {\bibfield  {journal} {\bibinfo
  {journal} {Phys. Chem. Chem. Phys.}\ }\textbf {\bibinfo {volume} {21}},\
  \bibinfo {pages} {17640} (\bibinfo {year} {2019})}\BibitemShut {NoStop}%
\bibitem [{\citenamefont {Gopinadhan}\ \emph {et~al.}(2019)\citenamefont
  {Gopinadhan}, \citenamefont {Hu}, \citenamefont {Esfandiar}, \citenamefont
  {Lozada-Hidalgo}, \citenamefont {Wang}, \citenamefont {Yang}, \citenamefont
  {Tyurnina}, \citenamefont {Keerthi}, \citenamefont {Radha},\ and\
  \citenamefont {Geim}}]{Geim2019}%
  \BibitemOpen
  \bibfield  {author} {\bibinfo {author} {\bibfnamefont {K.}~\bibnamefont
  {Gopinadhan}}, \bibinfo {author} {\bibfnamefont {S.}~\bibnamefont {Hu}},
  \bibinfo {author} {\bibfnamefont {A.}~\bibnamefont {Esfandiar}}, \bibinfo
  {author} {\bibfnamefont {M.}~\bibnamefont {Lozada-Hidalgo}}, \bibinfo
  {author} {\bibfnamefont {F.~C.}\ \bibnamefont {Wang}}, \bibinfo {author}
  {\bibfnamefont {Q.}~\bibnamefont {Yang}}, \bibinfo {author} {\bibfnamefont
  {A.~V.}\ \bibnamefont {Tyurnina}}, \bibinfo {author} {\bibfnamefont
  {A.}~\bibnamefont {Keerthi}}, \bibinfo {author} {\bibfnamefont
  {B.}~\bibnamefont {Radha}},\ and\ \bibinfo {author} {\bibfnamefont {A.~K.}\
  \bibnamefont {Geim}},\ }\bibfield  {title} {\bibinfo {title} {Complete steric
  exclusion of ions and proton transport through confined monolayer water},\
  }\href {https://doi.org/10.1126/science.aau6771} {\bibfield  {journal}
  {\bibinfo  {journal} {Science}\ }\textbf {\bibinfo {volume} {363}},\ \bibinfo
  {pages} {145} (\bibinfo {year} {2019})},\ \Eprint
  {https://arxiv.org/abs/https://www.science.org/doi/pdf/10.1126/science.aau6771}
  {https://www.science.org/doi/pdf/10.1126/science.aau6771} \BibitemShut
  {NoStop}%
\bibitem [{\citenamefont {Nair}\ \emph {et~al.}(2012)\citenamefont {Nair},
  \citenamefont {Wu}, \citenamefont {Jayaram}, \citenamefont {Grigorieva},\
  and\ \citenamefont {Geim}}]{Geim2012}%
  \BibitemOpen
  \bibfield  {author} {\bibinfo {author} {\bibfnamefont {R.~R.}\ \bibnamefont
  {Nair}}, \bibinfo {author} {\bibfnamefont {H.~A.}\ \bibnamefont {Wu}},
  \bibinfo {author} {\bibfnamefont {P.~N.}\ \bibnamefont {Jayaram}}, \bibinfo
  {author} {\bibfnamefont {I.~V.}\ \bibnamefont {Grigorieva}},\ and\ \bibinfo
  {author} {\bibfnamefont {A.~K.}\ \bibnamefont {Geim}},\ }\bibfield  {title}
  {\bibinfo {title} {Unimpeded permeation of water through helium-leak–tight
  graphene-based membranes},\ }\href {https://doi.org/10.1126/science.1211694}
  {\bibfield  {journal} {\bibinfo  {journal} {Science}\ }\textbf {\bibinfo
  {volume} {335}},\ \bibinfo {pages} {442} (\bibinfo {year}
  {2012})}\BibitemShut {NoStop}%
\bibitem [{\citenamefont {Xie}\ \emph {et~al.}(2018)\citenamefont {Xie},
  \citenamefont {Alibakhshi}, \citenamefont {Jiao}, \citenamefont {Xu},
  \citenamefont {Hempel}, \citenamefont {Kong}, \citenamefont {Park},\ and\
  \citenamefont {Duan}}]{FastWaterTransport}%
  \BibitemOpen
  \bibfield  {author} {\bibinfo {author} {\bibfnamefont {Q.}~\bibnamefont
  {Xie}}, \bibinfo {author} {\bibfnamefont {M.~A.}\ \bibnamefont {Alibakhshi}},
  \bibinfo {author} {\bibfnamefont {S.}~\bibnamefont {Jiao}}, \bibinfo {author}
  {\bibfnamefont {Z.}~\bibnamefont {Xu}}, \bibinfo {author} {\bibfnamefont
  {M.}~\bibnamefont {Hempel}}, \bibinfo {author} {\bibfnamefont
  {J.}~\bibnamefont {Kong}}, \bibinfo {author} {\bibfnamefont {H.~G.}\
  \bibnamefont {Park}},\ and\ \bibinfo {author} {\bibfnamefont
  {C.}~\bibnamefont {Duan}},\ }\bibfield  {title} {\bibinfo {title} {Fast water
  transport in graphene nanofluidic channels},\ }\href@noop {} {\bibfield
  {journal} {\bibinfo  {journal} {Nature Nanotechnology}\ }\textbf {\bibinfo
  {volume} {13}},\ \bibinfo {pages} {238} (\bibinfo {year} {2018})}\BibitemShut
  {NoStop}%
\bibitem [{\citenamefont {Wen}\ \emph {et~al.}(2022)\citenamefont {Wen},
  \citenamefont {Foller}, \citenamefont {Jin}, \citenamefont {Musso},
  \citenamefont {Kumar},\ and\ \citenamefont
  {Joshi}}]{UnderstandingWaterTransport}%
  \BibitemOpen
  \bibfield  {author} {\bibinfo {author} {\bibfnamefont {X.}~\bibnamefont
  {Wen}}, \bibinfo {author} {\bibfnamefont {T.}~\bibnamefont {Foller}},
  \bibinfo {author} {\bibfnamefont {X.}~\bibnamefont {Jin}}, \bibinfo {author}
  {\bibfnamefont {T.}~\bibnamefont {Musso}}, \bibinfo {author} {\bibfnamefont
  {P.}~\bibnamefont {Kumar}},\ and\ \bibinfo {author} {\bibfnamefont
  {R.}~\bibnamefont {Joshi}},\ }\bibfield  {title} {\bibinfo {title}
  {Understanding water transport through graphene-based nanochannels via
  experimental control of slip length},\ }\href@noop {} {\bibfield  {journal}
  {\bibinfo  {journal} {Nature Communications}\ }\textbf {\bibinfo {volume}
  {13}},\ \bibinfo {pages} {5690} (\bibinfo {year} {2022})}\BibitemShut
  {NoStop}%
\bibitem [{\citenamefont {Katsnelson}(2012)}]{KatsnelsonBook2012}%
  \BibitemOpen
  \bibfield  {author} {\bibinfo {author} {\bibfnamefont {M.~I.}\ \bibnamefont
  {Katsnelson}},\ }\href@noop {} {\emph {\bibinfo {title} {Graphene: Carbon in
  Two Dimensions}}}\ (\bibinfo  {publisher} {Cambridge University Press},\
  \bibinfo {year} {2012})\BibitemShut {NoStop}%
\bibitem [{\citenamefont {Wallbank}\ \emph {et~al.}(2019)\citenamefont
  {Wallbank}, \citenamefont {Krishna~Kumar}, \citenamefont {Holwill},
  \citenamefont {Wang}, \citenamefont {Auton}, \citenamefont {Birkbeck},
  \citenamefont {Mishchenko}, \citenamefont {Ponomarenko}, \citenamefont
  {Watanabe}, \citenamefont {Taniguchi}, \citenamefont {Novoselov},
  \citenamefont {Aleiner}, \citenamefont {Geim},\ and\ \citenamefont
  {Fal'ko}}]{Umklapp_2019}%
  \BibitemOpen
  \bibfield  {author} {\bibinfo {author} {\bibfnamefont {J.~R.}\ \bibnamefont
  {Wallbank}}, \bibinfo {author} {\bibfnamefont {R.}~\bibnamefont
  {Krishna~Kumar}}, \bibinfo {author} {\bibfnamefont {M.}~\bibnamefont
  {Holwill}}, \bibinfo {author} {\bibfnamefont {Z.}~\bibnamefont {Wang}},
  \bibinfo {author} {\bibfnamefont {G.~H.}\ \bibnamefont {Auton}}, \bibinfo
  {author} {\bibfnamefont {J.}~\bibnamefont {Birkbeck}}, \bibinfo {author}
  {\bibfnamefont {A.}~\bibnamefont {Mishchenko}}, \bibinfo {author}
  {\bibfnamefont {L.~A.}\ \bibnamefont {Ponomarenko}}, \bibinfo {author}
  {\bibfnamefont {K.}~\bibnamefont {Watanabe}}, \bibinfo {author}
  {\bibfnamefont {T.}~\bibnamefont {Taniguchi}}, \bibinfo {author}
  {\bibfnamefont {K.~S.}\ \bibnamefont {Novoselov}}, \bibinfo {author}
  {\bibfnamefont {I.~L.}\ \bibnamefont {Aleiner}}, \bibinfo {author}
  {\bibfnamefont {A.~K.}\ \bibnamefont {Geim}},\ and\ \bibinfo {author}
  {\bibfnamefont {V.~I.}\ \bibnamefont {Fal'ko}},\ }\bibfield  {title}
  {\bibinfo {title} {Excess resistivity in graphene superlattices caused by
  umklapp electron--electron scattering},\ }\href
  {https://doi.org/10.1038/s41567-018-0278-6} {\bibfield  {journal} {\bibinfo
  {journal} {Nature Physics}\ }\textbf {\bibinfo {volume} {15}},\ \bibinfo
  {pages} {32} (\bibinfo {year} {2019})}\BibitemShut {NoStop}%
\bibitem [{\citenamefont {Parmenter}(1953)}]{Parmenter53}%
  \BibitemOpen
  \bibfield  {author} {\bibinfo {author} {\bibfnamefont {R.~H.}\ \bibnamefont
  {Parmenter}},\ }\bibfield  {title} {\bibinfo {title} {The acousto-electric
  effect},\ }\href {https://doi.org/10.1103/PhysRev.89.990} {\bibfield
  {journal} {\bibinfo  {journal} {Phys. Rev.}\ }\textbf {\bibinfo {volume}
  {89}},\ \bibinfo {pages} {990} (\bibinfo {year} {1953})}\BibitemShut
  {NoStop}%
\bibitem [{\citenamefont {Eckstein}(1964)}]{Eckstein64}%
  \BibitemOpen
  \bibfield  {author} {\bibinfo {author} {\bibfnamefont {S.~G.}\ \bibnamefont
  {Eckstein}},\ }\bibfield  {title} {\bibinfo {title} {{Acoustoelectric
  Effect}},\ }\href {https://doi.org/10.1063/1.1713826} {\bibfield  {journal}
  {\bibinfo  {journal} {Journal of Applied Physics}\ }\textbf {\bibinfo
  {volume} {35}},\ \bibinfo {pages} {2702} (\bibinfo {year}
  {1964})}\BibitemShut {NoStop}%
\bibitem [{\citenamefont {Many}\ and\ \citenamefont
  {Balberg}(1969)}]{Many1969}%
  \BibitemOpen
  \bibfield  {author} {\bibinfo {author} {\bibfnamefont {A.}~\bibnamefont
  {Many}}\ and\ \bibinfo {author} {\bibfnamefont {I.}~\bibnamefont {Balberg}},\
  }\bibinfo {title} {The acoustoelectric effect},\ in\ \href
  {https://doi.org/10.1007/978-1-4899-6537-0_29} {\emph {\bibinfo {booktitle}
  {Electronic Structures in Solids: Lectures presented at the Second Chania
  Conference, held in Chania, Crete, June 30--July 14, 1968}}},\ \bibinfo
  {editor} {edited by\ \bibinfo {editor} {\bibfnamefont {E.~D.}\ \bibnamefont
  {Haidemenakis}}}\ (\bibinfo  {publisher} {Springer US},\ \bibinfo {address}
  {Boston, MA},\ \bibinfo {year} {1969})\ pp.\ \bibinfo {pages}
  {385--416}\BibitemShut {NoStop}%
\bibitem [{\citenamefont {Fal'ko}\ \emph {et~al.}(1993)\citenamefont {Fal'ko},
  \citenamefont {Meshkov},\ and\ \citenamefont {Iordanskii}}]{Falko93}%
  \BibitemOpen
  \bibfield  {author} {\bibinfo {author} {\bibfnamefont {V.~I.}\ \bibnamefont
  {Fal'ko}}, \bibinfo {author} {\bibfnamefont {S.~V.}\ \bibnamefont
  {Meshkov}},\ and\ \bibinfo {author} {\bibfnamefont {S.~V.}\ \bibnamefont
  {Iordanskii}},\ }\bibfield  {title} {\bibinfo {title} {Acoustoelectric drag
  effect in the two-dimensional electron gas at strong magnetic field},\ }\href
  {https://doi.org/10.1103/PhysRevB.47.9910} {\bibfield  {journal} {\bibinfo
  {journal} {Phys. Rev. B}\ }\textbf {\bibinfo {volume} {47}},\ \bibinfo
  {pages} {9910} (\bibinfo {year} {1993})}\BibitemShut {NoStop}%
\bibitem [{\citenamefont {Kalameitsev}\ \emph {et~al.}(2019)\citenamefont
  {Kalameitsev}, \citenamefont {Kovalev},\ and\ \citenamefont
  {Savenko}}]{ValleyDrag2019}%
  \BibitemOpen
  \bibfield  {author} {\bibinfo {author} {\bibfnamefont {A.~V.}\ \bibnamefont
  {Kalameitsev}}, \bibinfo {author} {\bibfnamefont {V.~M.}\ \bibnamefont
  {Kovalev}},\ and\ \bibinfo {author} {\bibfnamefont {I.~G.}\ \bibnamefont
  {Savenko}},\ }\bibfield  {title} {\bibinfo {title} {Valley acoustoelectric
  effect},\ }\href {https://doi.org/10.1103/PhysRevLett.122.256801} {\bibfield
  {journal} {\bibinfo  {journal} {Phys. Rev. Lett.}\ }\textbf {\bibinfo
  {volume} {122}},\ \bibinfo {pages} {256801} (\bibinfo {year}
  {2019})}\BibitemShut {NoStop}%
\bibitem [{\citenamefont {Lapa}\ \emph {et~al.}(2020)\citenamefont {Lapa},
  \citenamefont {Kassabian}, \citenamefont {Torres}, \citenamefont {Salev},
  \citenamefont {Lee}, \citenamefont {del Valle},\ and\ \citenamefont
  {Schuller}}]{Lapa2020}%
  \BibitemOpen
  \bibfield  {author} {\bibinfo {author} {\bibfnamefont {P.~N.}\ \bibnamefont
  {Lapa}}, \bibinfo {author} {\bibfnamefont {G.}~\bibnamefont {Kassabian}},
  \bibinfo {author} {\bibfnamefont {F.}~\bibnamefont {Torres}}, \bibinfo
  {author} {\bibfnamefont {P.}~\bibnamefont {Salev}}, \bibinfo {author}
  {\bibfnamefont {M.-H.}\ \bibnamefont {Lee}}, \bibinfo {author} {\bibfnamefont
  {J.}~\bibnamefont {del Valle}},\ and\ \bibinfo {author} {\bibfnamefont
  {I.~K.}\ \bibnamefont {Schuller}},\ }\bibfield  {title} {\bibinfo {title}
  {{Acoustoelectric drag current in vanadium oxide films}},\ }\href
  {https://doi.org/10.1063/5.0015215} {\bibfield  {journal} {\bibinfo
  {journal} {Journal of Applied Physics}\ }\textbf {\bibinfo {volume} {128}},\
  \bibinfo {pages} {155104} (\bibinfo {year} {2020})}\BibitemShut {NoStop}%
\bibitem [{\citenamefont {Castro~Neto}\ \emph {et~al.}(2009)\citenamefont
  {Castro~Neto}, \citenamefont {Guinea}, \citenamefont {Peres}, \citenamefont
  {Novoselov},\ and\ \citenamefont {Geim}}]{RevModPhys.81.109}%
  \BibitemOpen
  \bibfield  {author} {\bibinfo {author} {\bibfnamefont {A.~H.}\ \bibnamefont
  {Castro~Neto}}, \bibinfo {author} {\bibfnamefont {F.}~\bibnamefont {Guinea}},
  \bibinfo {author} {\bibfnamefont {N.~M.~R.}\ \bibnamefont {Peres}}, \bibinfo
  {author} {\bibfnamefont {K.~S.}\ \bibnamefont {Novoselov}},\ and\ \bibinfo
  {author} {\bibfnamefont {A.~K.}\ \bibnamefont {Geim}},\ }\bibfield  {title}
  {\bibinfo {title} {The electronic properties of graphene},\ }\href
  {https://doi.org/10.1103/RevModPhys.81.109} {\bibfield  {journal} {\bibinfo
  {journal} {Rev. Mod. Phys.}\ }\textbf {\bibinfo {volume} {81}},\ \bibinfo
  {pages} {109} (\bibinfo {year} {2009})}\BibitemShut {NoStop}%
\bibitem [{\citenamefont {Rosenzweig}\ \emph {et~al.}(2019)\citenamefont
  {Rosenzweig}, \citenamefont {Karakachian}, \citenamefont {Link},
  \citenamefont {K\"uster},\ and\ \citenamefont
  {Starke}}]{Doping_PhysRevB.100.035445}%
  \BibitemOpen
  \bibfield  {author} {\bibinfo {author} {\bibfnamefont {P.}~\bibnamefont
  {Rosenzweig}}, \bibinfo {author} {\bibfnamefont {H.}~\bibnamefont
  {Karakachian}}, \bibinfo {author} {\bibfnamefont {S.}~\bibnamefont {Link}},
  \bibinfo {author} {\bibfnamefont {K.}~\bibnamefont {K\"uster}},\ and\
  \bibinfo {author} {\bibfnamefont {U.}~\bibnamefont {Starke}},\ }\bibfield
  {title} {\bibinfo {title} {Tuning the doping level of graphene in the
  vicinity of the van hove singularity via ytterbium intercalation},\ }\href
  {https://doi.org/10.1103/PhysRevB.100.035445} {\bibfield  {journal} {\bibinfo
   {journal} {Phys. Rev. B}\ }\textbf {\bibinfo {volume} {100}},\ \bibinfo
  {pages} {035445} (\bibinfo {year} {2019})}\BibitemShut {NoStop}%
\bibitem [{\citenamefont {Rosenzweig}\ \emph {et~al.}(2020)\citenamefont
  {Rosenzweig}, \citenamefont {Karakachian}, \citenamefont {Marchenko},
  \citenamefont {K\"uster},\ and\ \citenamefont
  {Starke}}]{Doping_PhysRevLett.125.176403}%
  \BibitemOpen
  \bibfield  {author} {\bibinfo {author} {\bibfnamefont {P.}~\bibnamefont
  {Rosenzweig}}, \bibinfo {author} {\bibfnamefont {H.}~\bibnamefont
  {Karakachian}}, \bibinfo {author} {\bibfnamefont {D.}~\bibnamefont
  {Marchenko}}, \bibinfo {author} {\bibfnamefont {K.}~\bibnamefont
  {K\"uster}},\ and\ \bibinfo {author} {\bibfnamefont {U.}~\bibnamefont
  {Starke}},\ }\bibfield  {title} {\bibinfo {title} {Overdoping graphene beyond
  the van hove singularity},\ }\href
  {https://doi.org/10.1103/PhysRevLett.125.176403} {\bibfield  {journal}
  {\bibinfo  {journal} {Phys. Rev. Lett.}\ }\textbf {\bibinfo {volume} {125}},\
  \bibinfo {pages} {176403} (\bibinfo {year} {2020})}\BibitemShut {NoStop}%
\bibitem [{\citenamefont {Di~Ventra}(2008)}]{DiVentra_2008}%
  \BibitemOpen
  \bibfield  {author} {\bibinfo {author} {\bibfnamefont {M.}~\bibnamefont
  {Di~Ventra}},\ }\href@noop {} {\emph {\bibinfo {title} {Electrical Transport
  in Nanoscale Systems}}}\ (\bibinfo  {publisher} {Cambridge University
  Press},\ \bibinfo {year} {2008})\BibitemShut {NoStop}%
\bibitem [{\citenamefont {Kamenev}(2023)}]{kamenev2023field}%
  \BibitemOpen
  \bibfield  {author} {\bibinfo {author} {\bibfnamefont {A.}~\bibnamefont
  {Kamenev}},\ }\href {https://books.google.de/books?id=y9WgEAAAQBAJ} {\emph
  {\bibinfo {title} {Field Theory of Non-Equilibrium Systems}}}\ (\bibinfo
  {publisher} {Cambridge University Press},\ \bibinfo {year}
  {2023})\BibitemShut {NoStop}%
\bibitem [{\citenamefont {Sieberer}\ \emph {et~al.}(2016)\citenamefont
  {Sieberer}, \citenamefont {Buchhold},\ and\ \citenamefont
  {Diehl}}]{Sieberer_2016}%
  \BibitemOpen
  \bibfield  {author} {\bibinfo {author} {\bibfnamefont {L.~M.}\ \bibnamefont
  {Sieberer}}, \bibinfo {author} {\bibfnamefont {M.}~\bibnamefont {Buchhold}},\
  and\ \bibinfo {author} {\bibfnamefont {S.}~\bibnamefont {Diehl}},\ }\bibfield
   {title} {\bibinfo {title} {Keldysh field theory for driven open quantum
  systems},\ }\href {https://doi.org/10.1088/0034-4885/79/9/096001} {\bibfield
  {journal} {\bibinfo  {journal} {Reports on Progress in Physics}\ }\textbf
  {\bibinfo {volume} {79}},\ \bibinfo {pages} {096001} (\bibinfo {year}
  {2016})}\BibitemShut {NoStop}%
\bibitem [{\citenamefont {Fr\"a\ss{}dorf}\ and\ \citenamefont
  {Mosig}(2017)}]{Keldysh_FRG_PhysRevB.95.125412}%
  \BibitemOpen
  \bibfield  {author} {\bibinfo {author} {\bibfnamefont {C.}~\bibnamefont
  {Fr\"a\ss{}dorf}}\ and\ \bibinfo {author} {\bibfnamefont {J.~E.~M.}\
  \bibnamefont {Mosig}},\ }\bibfield  {title} {\bibinfo {title} {Keldysh
  functional renormalization group for electronic properties of graphene},\
  }\href {https://doi.org/10.1103/PhysRevB.95.125412} {\bibfield  {journal}
  {\bibinfo  {journal} {Phys. Rev. B}\ }\textbf {\bibinfo {volume} {95}},\
  \bibinfo {pages} {125412} (\bibinfo {year} {2017})}\BibitemShut {NoStop}%
\bibitem [{\citenamefont {Oka}\ and\ \citenamefont
  {Aoki}(2009)}]{Aoki_PhysRevB.79.081406}%
  \BibitemOpen
  \bibfield  {author} {\bibinfo {author} {\bibfnamefont {T.}~\bibnamefont
  {Oka}}\ and\ \bibinfo {author} {\bibfnamefont {H.}~\bibnamefont {Aoki}},\
  }\bibfield  {title} {\bibinfo {title} {Photovoltaic hall effect in
  graphene},\ }\href {https://doi.org/10.1103/PhysRevB.79.081406} {\bibfield
  {journal} {\bibinfo  {journal} {Phys. Rev. B}\ }\textbf {\bibinfo {volume}
  {79}},\ \bibinfo {pages} {081406} (\bibinfo {year} {2009})}\BibitemShut
  {NoStop}%
\bibitem [{\citenamefont {Chudnovskiy}()}]{Supplement}%
  \BibitemOpen
  \bibfield  {author} {\bibinfo {author} {\bibfnamefont {A.~L.}\ \bibnamefont
  {Chudnovskiy}},\ }\bibfield  {title} {\bibinfo {title} {Supplemental
  material},\ }\href@noop {} {\bibinfo  {journal} {see Supplemental material at
  [ ] for detailed calculatons}\ }\BibitemShut {NoStop}%
\bibitem [{\citenamefont {Deyo}\ \emph {et~al.}(2009)\citenamefont {Deyo},
  \citenamefont {Golub}, \citenamefont {Ivchenko},\ and\ \citenamefont
  {Spivak}}]{deyo2009}%
  \BibitemOpen
\bibfield  {journal} {  }\bibfield  {author} {\bibinfo {author} {\bibfnamefont
  {E.}~\bibnamefont {Deyo}}, \bibinfo {author} {\bibfnamefont {L.~E.}\
  \bibnamefont {Golub}}, \bibinfo {author} {\bibfnamefont {E.~L.}\ \bibnamefont
  {Ivchenko}},\ and\ \bibinfo {author} {\bibfnamefont {B.}~\bibnamefont
  {Spivak}},\ }\href {https://arxiv.org/abs/0904.1917} {\bibinfo {title}
  {Semiclassical theory of the photogalvanic effect in non-centrosymmetric
  systems}} (\bibinfo {year} {2009}),\ \Eprint
  {https://arxiv.org/abs/0904.1917} {arXiv:0904.1917} \BibitemShut {NoStop}%
\bibitem [{\citenamefont {Sodemann}\ and\ \citenamefont
  {Fu}(2015)}]{Sodemann2015}%
  \BibitemOpen
  \bibfield  {author} {\bibinfo {author} {\bibfnamefont {I.}~\bibnamefont
  {Sodemann}}\ and\ \bibinfo {author} {\bibfnamefont {L.}~\bibnamefont {Fu}},\
  }\bibfield  {title} {\bibinfo {title} {Quantum nonlinear hall effect induced
  by berry curvature dipole in time-reversal invariant materials},\ }\href
  {https://doi.org/10.1103/PhysRevLett.115.216806} {\bibfield  {journal}
  {\bibinfo  {journal} {Phys. Rev. Lett.}\ }\textbf {\bibinfo {volume} {115}},\
  \bibinfo {pages} {216806} (\bibinfo {year} {2015})}\BibitemShut {NoStop}%
\end{thebibliography}%

\newpage 
\begin{center}
\section*{ Supplemental Material}
\end{center}

In this Supplemental Material we provide detailed derivations of the formulas presented in the main text of the paper.  To keep the presentation self-consistent, we repeat some expressions given  in the main text. \\ 

\section{Important length scales} 
Let us  list the  important length scales in graphene that will be used for further estimations.  
Graphene lattice constant $a\approx 1.42$ \AA.  Lattice vectors in graphene: 
\begin{equation}
	{\bf a}_1=\frac{a}{2}(3, \sqrt{3}), \, \, \, {\bf a}_2=\frac{a}{2}(3, -\sqrt{3})
\end{equation}

Vectors connecting the nearest neighbour atoms in graphene 
\begin{equation}
	{\bm \delta}_1=\frac{a}{2}(1, \sqrt{3}), \, \, \, {\bm \delta}_2=\frac{a}{2}(1, - \sqrt{3}), {\bm \delta}_3=-a(1, 0). 
\end{equation}

Corresponding reciprocal-lattice vectors 
\begin{equation}
	{\bf b}_1=\frac{2\pi}{3a} (1, \sqrt{3}), \, \, \, 	{\bf b}_2=\frac{2\pi}{3a} (1, -\sqrt{3})
\end{equation}
The $K$-points 
\begin{equation}
	{\bf K'}=\frac{2\pi}{3a} (1, 1/\sqrt{3}), \, \, \, {\bf K}=\frac{2\pi}{3a} (1, -1/\sqrt{3}). 
\end{equation}
There are also equivalent $K$ and $K'$ points given by the vectors 
\begin{equation}
	{\bf K'}<=>{\bf K'}-{\bf b}_1<=>{\bf K'}-{\bf b}_1-{\bf b}_2, \, \, \, 	{\bf K}<=>{\bf K}-{\bf b}_2<=>{\bf K}-{\bf b}_1-{\bf b}_2. 
\end{equation}
The lattice constant of the square ice crystal  $a_0\sim 2.8$ \AA. Furthermore we put $\hbar=e=1$ in all intermediate formulas.

\section{Operator of electric current density in the tight-binding model of graphene}
Since spin plays no role in further considerations, we suppress spin-indexes here and in what follows.  The tight binding Hamiltonian  of graphene (here the sublattices $A, B$ correspond to pseudospin) is given by 
\begin{equation}
	H=i t\sum_ i \sum_{j=1}^3(\psi^+_A({\bf r}_i) \psi_B({\bf r}_i+{\bf \delta}_j)  -  \psi_B^+({\bf r}_i+{\bf \delta}_j)  \psi_A({\bf r}_i))=\sum_{\bf k} (\psi_A^+({\bf k}), \psi_B^+({\bf k}))\left(\begin{array}{cc} 
		0 & i t\sum_{\langle j \rangle} e^{i{\bf k}{\bm \delta}_j} \\ 
		- i t\sum_{\langle j \rangle} e^{-i{\bf k}{\bm \delta}_j} & 0 
	\end{array} \right) 
	\left( \begin{array}{c} \psi_A({\bf k}) \\ \psi_B({\bf k}) \end{array} \right). 
\end{equation}
Here $\psi^+_{A,B}({\bf r}_i), \psi_{A,B}({\bf r}_i) $ denote fermion creation and annihilation operators at the site $i$ of $A$ and $B$ sublattices respectively, and the sum over $j$ runs over the nearest neighbour cites.  
External vector potential induces phase shifts on links $t_{ij}\rightarrow t_{ij} e^{i\phi_{ij}}$, where $\phi_{ij}= {\bf A}_i \cdot {\bm\delta}_j$ (the scalar product of the vector potential at the site $i$ and the vector to the next-nearest neighbour $j$).   In the (real-time) action, the coupling to the external vector potential is described by the following term 
\begin{equation}
	S_A=-\int_t\sum_{ i} \sum_{j=1}^3 (\bar{\psi}_A({\bf r}_i), \bar{\psi}_{B}({\bf r}_i+{\bm \delta}_j))\left(\begin{array}{cc} 
		0 & i  t e^{i\phi_{ij}} \\ 
		- i  t e^{-i\phi_{ij}} & 0 \end{array}\right) 
	\left(\begin{array}{c} \psi_A({\bf r}_i)\\ \psi_{B}({\bf r}_i+{\bm \delta}_j)  \end{array}\right).  
\end{equation}
Transforming to the ${\bf k}$-space according to 
\begin{equation}
	\psi_A({\bf r})=\int_{\bf k} \psi_A({\bf k}) e^{i{\bf k}{\bf r}}, \, \, \, \psi_B({\bf r}+{\bm \delta}_3)=\int_{\bf k} \psi_B({\bf k}) e^{i{\bf k}{\bf r}}, 
	\label{FT}
\end{equation}
we obtain 
\begin{equation}
	S_A=-\int_t \sum_{ i} \sum_{j=1}^3 \int_{\bf k,k'} \left(\bar{\psi}_A({\bf k'}), \bar{\psi}_B({\bf k'}) e^{-i{\bf k'}({\bm{\delta}_j}-{\bm{\delta}_3})} \right)e^{-i{\bf k'}{\bf r}_i} \left(\begin{array}{cc} 
		0 & i t e^{i\phi_{ij}} \\ 
		- i t e^{-i\phi_{ij}} & 0 \end{array}\right) e^{i{\bf k}{\bf r}_i}
	\left(\begin{array}{c} \psi_A({\bf k}) \\ \psi_B({\bf k}) e^{i{\bf k}(\bm{\delta}_j- \bm{\delta}_3)} \end{array}\right).  
\end{equation}
The expression for the operator of the current density is obtained by the variation of the action with respect to the vector potential. 
\begin{equation}
	{\bf j}({\bf r}_i)=\frac{\delta S}{\delta {\bf A}({\bf r}_i)}= \sum_{\langle j \rangle} {\bm \delta}_j \frac{\delta S}{\delta\phi_{ij}}\bigg{|}_{\phi_{ij}=0} =
	\int_{{\bf k},{\bf k'}} \sum_{ j=1}^3 e^{i ({\bf k}-{\bf k'}) {\bf r}_i} \left(\bar{\psi}_A({\bf k'}), \bar{\psi}_B({\bf k'})\right) 
	\left(\begin{array}{cc} 
		0 &  - {\bm \delta}_j  t   e^{ i{\bf k}({\bm \delta}_j -{\bm \delta}_3) } \\ 
		- {\bm \delta}_j  t   e^{-i {\bf k'} ({\bm\delta}_j -{\bm\delta}_3) } & 0 \end{array}\right) \left(\begin{array}{c} \psi_A({\bf k}) \\ \psi_B({\bf k})  \end{array}\right). 
\end{equation}

\subsection{Expansion of the current density operator for the states close to $K$ and $K'$ points.  }
Assume ${\bf k}={\bf K}+{\bf p}$, ${\bf k'}={\bf K'}+{\bf p'}$, ($|{\bf p}|, |{\bf p'}| \ll K, K'$). Then the current density operator can be represented as a sum of intra- and inter-valley components 
\begin{equation}
	{\bf j}({\bf r}_i)= 	{\bf j}_{KK}({\bf r}_i)+	{\bf j}_{K'K'}({\bf r}_i)+ 	{\bf j}_{KK'}({\bf r}_i)+	{\bf j}_{K'K}({\bf r}_i), 
\end{equation}
where for instance 
\begin{equation}
	{\bf j}_{K'K}({\bf r})= -t\sum_{j=1}^3
	\int_{{\bf p},{\bf p'}} e^{i ({\bf K}-{\bf K'}+{\bf p}-{\bf p'}) {\bf r}} \left(\bar{\psi}_A({{\bf K'}+\bf p'}), \bar{\psi}_B({{\bf K'}+\bf p'}) \right) 
	\left(\begin{array}{cc} 
		0 &    {\bm \delta}_j  e^{ i({\bf K}+{\bf p})({\bm \delta}_j -{\bm\delta}_3) } \\ 
		{\bm \delta}_j  e^{-i ({\bf K'}+{\bf p'})( {\bm\delta}_j-{\bm\delta}_3)  } & 0 \end{array}\right) \left(\begin{array}{c} \psi_A({{\bf K}+\bf p}) \\ \psi_B({{\bf K}+\bf p})  \end{array}\right). 
\end{equation}
Let us introduce  the pseudospin spinors for the states close to the $K$-points  
$\Psi_{K}({\bf p})=(\psi_A({\bf K+p}), \psi_B({\bf K+p}))$,  $\Psi_{K'}({\bf p'})=(\psi_A({\bf K'+p'}), \psi_B({\bf K'+p'}))$. 
Performing the Fourier transform and denoting ${\bf q}={\bf p'}-{\bf p}$, we can write the Fourier component at the vector ${\bf K}-{\bf K'}-{\bf q}$ as 
\begin{eqnarray}
	\nonumber && 
	{\bf j}_{K'K}({\bf q})= 
	\int_{\bf p} \bar{\Psi}_{K'}({\bf p+q})
	\left(\begin{array}{cc} 
		0 &  - t \sum_{\langle j \rangle}  {\bm \delta}_j  e^{ i({\bf K}+{\bf p})({\bm \delta}_j-{\bm\delta}_3)  } \\ 
		-  t  \sum_{\langle j\rangle} {\bm \delta}_j  e^{-i ({\bf K'}+{\bf p}+{\bf q}) ({\bm\delta}_j-{\bm\delta}_3)  } & 0 \end{array}\right)  \Psi_K({\bf p} )  
\end{eqnarray}

Let us introduce the operators of partial current density in each valley, which are defined as follows 
\begin{equation}
	{\bf j}_K({\bf p})=- t \sum_{j=1}^3  {\bm \delta}_j  e^{ i({\bf K}+{\bf p})({\bm \delta}_j-{\bm\delta}_3)  }, \, \, \, \, 	{\bf j}_{K'}({\bf p})=- t \sum_{\langle j \rangle}  {\bm \delta}_j  e^{ i({\bf K'}+{\bf p})({\bm \delta}_j-{\bm\delta}_3)  }. 
	\label{jPartial}
\end{equation} 
Then the inter- and intra-valley components of the current density operator can be cast to the form 
\begin{equation}
	{\bf j}_{ab}({\bf q})= 
	\int_{\bf p} (\bar{\Psi}_a({\bf p+q}) 	{\bf J}_{ab}({\bf p}, {\bf p+q})
	\Psi_b ({\bf p}), 
\end{equation}
where 
\begin{equation}
	{\bf J}_{ab}({\bf p+q}, {\bf p})= 	\left(\begin{array}{cc} 
		0 &  {\bf j}_b({\bf p})  \\ 
		{\bf j}^*_a({\bf p}+{\bf q})  & 0 \end{array}\right).
	\label{Jab}
\end{equation}
Here the subscripts $a, b$ relate to the valley, taking the values $K, K'$. 

\section{Green functions near $K$ points in the Keldysh formalism} 
The Hamiltonian of graphene for the states close to the $K, K'$-points (${\bf K}=\frac{2\pi}{3a} (1, -1/\sqrt{3})$, ${\bf K'}=\frac{2\pi}{3a} (1, 1/\sqrt{3})$) reads 
\begin{eqnarray}
	&& 
	H_K=v_F\int_{{\bf p}}\Psi_{K}^+({\bf p}) (\sigma_x p_x+\sigma_y p_y)\Psi_{K}({\bf p}), \label{HK}  \\ 	
	&& 
	H_{K'}=v_F\int_{{\bf p}}\Psi_{K'}^+({\bf p}) ({\bf p}) (\sigma_x p_x-\sigma_y p_y)\Psi_{K'}({\bf p}),  \label{HK'} 
\end{eqnarray}
where $v_F=\frac{3}{2} t a$ is the Fermi velocity. 
The Green functions in the Keldysh formalism have a matrix structure given by 
\begin{equation}
	\hat{G}(\epsilon, {\bf p})=\left( 
	\begin{array}{cc} 
		G^R(\epsilon, {\bf p}) & 	G^K(\epsilon, {\bf p}) \\ 
		0 & 	G^A(\epsilon, {\bf p}) 
	\end{array}\right).
\end{equation}
Here the superscripts $R/A/K$ relate to the retarded, advanced, and Keldysh components of the Green function respectively. 
According to the Hamiltonian Eq. (\ref{HK}), close to the $K$-point,  the explicit form of the Green functions in equilibrium at temperature $T$ is given by 
\begin{eqnarray}
	&& 
	G^{R/A}_K(\epsilon, {\bf p})=\frac{\epsilon {\bf 1}_2+v_F {\bf p}\cdot \bm{\sigma}}{\epsilon^2-v_F^2p^2\pm io}= 
	\frac{1}{\epsilon^2-v_F^2p^2\pm io}
	\left(\begin{array}{cc}
		\epsilon & v_F(p_x-ip_y) \\ v_F(p_x+ip_y) & \epsilon \end{array} 	\right) \\ 
	&& 
	G^K_K(\epsilon, {\bf p})=\tanh\left(\frac{\epsilon}{2T}\right) \left(	G^{R}(\epsilon, {\bf p}) -  G^{A}(\epsilon, {\bf p}) \right).
\end{eqnarray}
In what follows we introduce complex momentum $p=p_x+ip_y$, denoting $\bar{p}=p_x-ip_y$, and $|p|=\sqrt{p_x^2+p_y^2}$. Then the expressions for the Green functions can be represented in the form that exhibits their pole structure explicitly 
\begin{equation}
	G_K^{R/A} (\epsilon, {\bf p})=\frac{1}{2}\left[\frac{1}{\epsilon-v_F|p|\pm io} + \frac{1}{\epsilon+v_F|p|\pm io}\right] 
	\left(\begin{array}{cc} 
		1 & v_F \bar{p}/\epsilon \\  v_F p/\epsilon & 1 \end{array}  
	\right),
\end{equation}
\begin{equation}
	G^K_K (\epsilon, {\bf p})=- i \pi  \tanh\left(\frac{\epsilon}{2T}\right) \left(\begin{array}{cc} 
		1 & v_F \bar{p}/\epsilon \\  v_F p/\epsilon & 1 \end{array}  
	\right) [\delta(\epsilon-v_F|p|)+ \delta(\epsilon+v_F|p|)]. 
\end{equation}

The Green functions for momenta close to the $K'$ point are determined by the Hamiltonian Eq. (\ref{HK'}). They  read 
\begin{equation}
	G^{R/A}_{K'}(\epsilon, {\bf p})=\frac{\epsilon {\bf 1}_2+v_F {\bf p}\cdot \bm{\sigma}^*}{\epsilon^2-v_F^2p^2\pm io}= 
	\frac{1}{\epsilon^2-v_F^2p^2\pm io}
	\left(\begin{array}{cc}
		\epsilon & v_F(p_x+ip_y) \\ v_F(p_x-ip_y) & \epsilon \end{array} 	\right), 
\end{equation}
or, in the form explicitly exhibiting the pole structure, 
\begin{equation}
	G_{K'}^{R/A} (\epsilon, {\bf p})=\frac{1}{2}\left[\frac{1}{\epsilon-v_F|p|\pm io} + \frac{1}{\epsilon+v_F|p|\pm io}\right] 
	\left(\begin{array}{cc} 
		1 & v_F p/\epsilon \\  v_F  \bar{p}/\epsilon & 1 \end{array}  
	\right),
\end{equation}
\begin{equation}
	G^K_{K'} (\epsilon, {\bf p})=- i \pi  \tanh\left(\frac{\epsilon}{2T}\right) \left(\begin{array}{cc} 
		1 & v_F p/\epsilon \\  v_F \bar{p}/\epsilon & 1 \end{array}  
	\right) [\delta(\epsilon-v_F|p|)+ \delta(\epsilon+v_F|p|)]. 
\end{equation}
Since the Green function in different valleys differ only  by the matrix structure in the pseudospin space, it is convenient to introduce the  following representation 
\begin{equation}
	\hat{G}_{\nu}^{R/A} (\epsilon, {\bf p})=G^{R/A} (\epsilon,{\bf p}) \hat{g}_{\nu}(\epsilon, {\bf p}) , \, \, \,  G_{\nu}^K (\epsilon, {\bf p})= G^K(\epsilon, {\bf p} )  \hat{g}_{\nu}(\epsilon, {\bf p}),  
	\label{NotationsGF}
\end{equation}
where 
\begin{eqnarray}
	&&  
	G^{R/A} (\epsilon,{\bf p}) =
	\frac{1}{2}\left[\frac{1}{\epsilon-v_F|p|\pm io} + \frac{1}{\epsilon+v_F|p|\pm io}\right],  \label{G_RA}\\ 
	&& 
	G^K (\epsilon, {\bf p})=  - i \pi  \tanh\left(\frac{\epsilon}{2T}\right) 
	[\delta(\epsilon-v_F|p|)+ \delta(\epsilon+v_F|p|)],  \label{G_K}
\end{eqnarray} 
denote the parts of the Green functions independent of their pseudospin structure, and 
\begin{eqnarray}	 
	&&  	  
	\hat{g}_{K'}(\epsilon, {\bf p})= 	\left(\begin{array}{cc} 
		1 & v_F p/\epsilon \\  v_F \bar{p}/\epsilon & 1 \end{array}  
	\right)=\sigma_0+ \frac{v_F}{\epsilon}(p_x\sigma_x-p_y\sigma_y),\\ 
	&& 
	\hat{g}_{K}(\epsilon, {\bf p})= 	\left(\begin{array}{cc} 
		1 & v_F \bar{p}/\epsilon \\  v_F p/\epsilon & 1 \end{array}  
	\right)=\sigma_0+ \frac{v_F}{\epsilon}(p_x\sigma_x+p_y\sigma_y),
\end{eqnarray}
capture the structure of Green functions in the pseudospin space.

\section{Projection on the conduction band}

Due to the necessity of finite doping, the main contribution to the drag current at low temperature is produced by transitions between the quantum states in the conduction band. This observation allows simplified calculation of the current by projection on the conduction band. To realize the projection,  we first employ the unitary rotation that diagonalizes the Hamiltonian in the pseudospin sector. Let $\phi_p$ be the polar angle of the wave vector ${\bf p}=(p_x, p_y)$. Then the Hamiltonian of the state with the wave vector ${\bf p}$ is diagonalized by the unitary transformation 
\begin{equation}
	H({\bf p})=\left(\begin{array}{cc}
		v_F |p| & 0 \\ 0 & -v_F |p|  
	\end{array}\right) = U_{\nu}^+(\phi_p) H_{\nu}({\bf p} ) U_{\nu}(\phi_p), 
\end{equation}
where 
\begin{equation}
	U_{K}({\bf p})=\frac{1}{\sqrt{2}}\left(\begin{array}{cc}
		e^{-i\phi_{\bf p}/2} & 	e^{-i\phi_{\bf p}/2} \\ 
		e^{i\phi_{\bf p}/2} & -e^{i\phi_{\bf p}/2}
	\end{array}\right), \, \, \, \,  	U_{K'}({\bf p})=\frac{1}{\sqrt{2}}\left(\begin{array}{cc}
		e^{i\phi_{\bf p}/2} & 	e^{i\phi_{\bf p}/2} \\ 
		e^{-i\phi_{\bf p}/2} & -e^{-i\phi_{\bf p}/2}
	\end{array}\right)
\end{equation}
Further projection on the conduction band is achieved by the application of the operator $\frac{1}{2}(1+\sigma_z)$, so that the total projection operator reads 
\begin{equation}
	\hat{P}^c_{\nu} ({\bf p}) =U_{\nu}({\bf p}) \left(\begin{array}{cc}
		1 & 	0 \\ 
		0& 0
	\end{array}\right)
\end{equation}
Explicitly, close to the points $K$ and $K'$,  the eigenstates of the conduction band are expressed through the eigenstates of  $A$ and $B$ sublattices as follows 
\begin{equation}
	|p\rangle_c^K=\frac{1}{\sqrt{2}} \left( e^{-i\phi_{\bf p}/2} |A\rangle + e^{i\phi_{\bf p}/2} |B\rangle\right), \, \, \, \, 
	|p\rangle_c^{K'}=\frac{1}{\sqrt{2}} \left( e^{i\phi_{\bf p}/2} |A\rangle + e^{-i\phi_{\bf p}/2} |B\rangle\right)
\end{equation}

Let us now consider the kinematic restriction on the wave vectors of the initial and final states due to the energy and quasi-momentum conservation in detail.  
Since the energy transfer by the scattering processes considered here is much smaller than the Fermi energy, we can approximately set the absolute value of the wave vectors for the initial and final states equal to the Fermi wave vector $p_F$.  
It turns out that this condition leaves only two possible choices for the wave vector of the initial state ${\bf p}$ (the wave vector of the final state is then fixed automatically to ${\bf p} +{\bf q}$) for the fixed transferred wave vector ${\bf q}$ (see Fig. \ref{fig:Angles}). Denote the polar angle of the wave vector $\bf q$ as $\alpha$ and the angle between vectors ${\bf q}$ and ${\bf p}$ as $\phi$. Then the two possible choices are 
\begin{figure}[t!]
	\centering
	\includegraphics[width=0.5\textwidth]{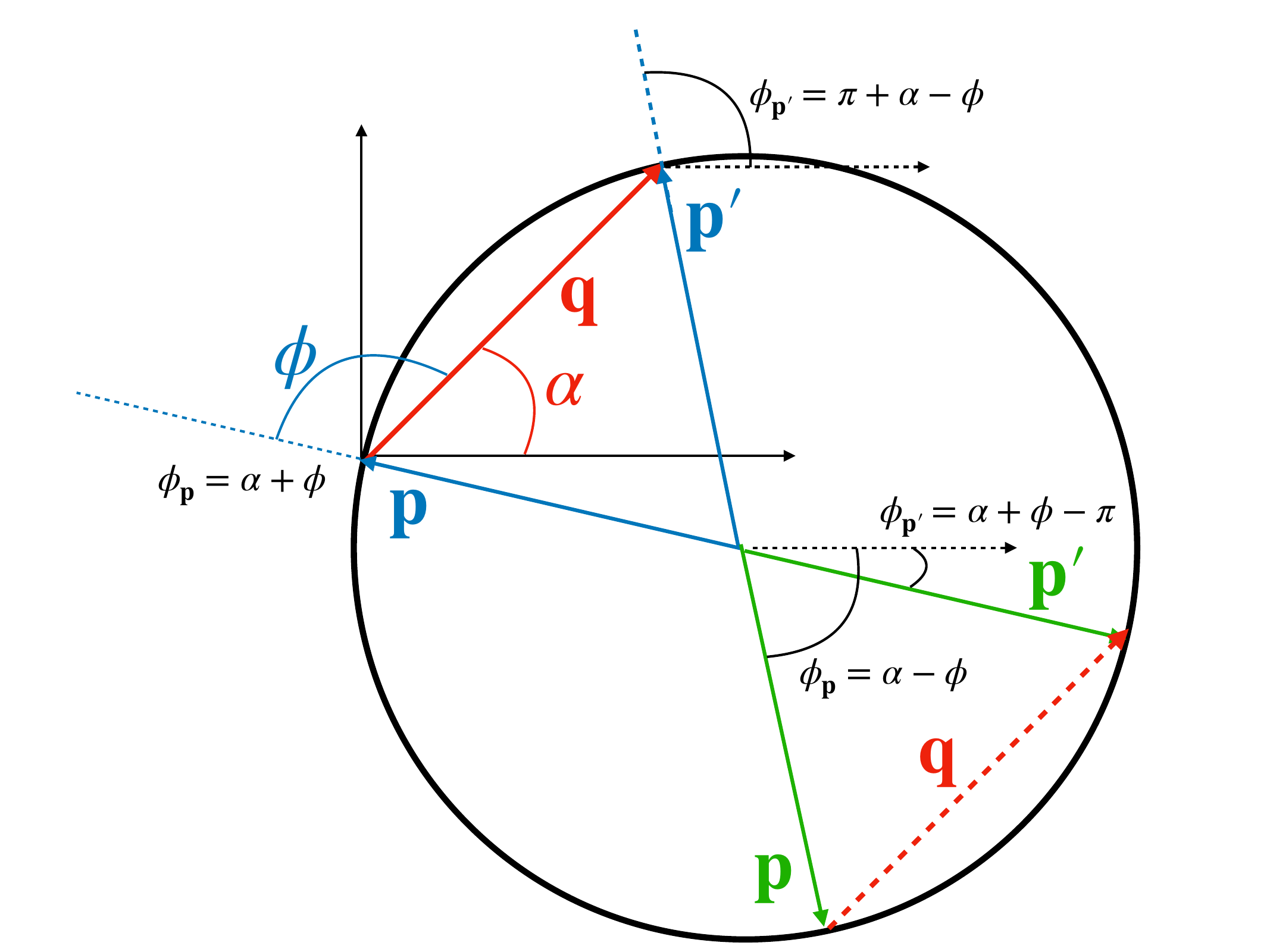}
	\caption{Possible directions of initial and finite wave vectors for a given transfer wave vector ${\bf q}$}
	\label{fig:Angles}
\end{figure} 
\begin{eqnarray}
	&& 
	{\bf i)}\, \, \, 	{\phi_{\bf p}}=\alpha+\phi, \, \, \, {\phi_{\bf p'}}=\pi+ \alpha-\phi,  \label{phiphi'_i} \\ 
	&& 
	{\bf ii)}\, \, \, 
	{\phi_{\bf p}}=\alpha-\phi, \, \, \, {\phi_{\bf p'}}= \alpha+\phi-\pi <=> \pi + \alpha+\phi. \label{phiphi'_ii}
\end{eqnarray}
It follows that all results for the case (ii) are obtained from (i) by change $\phi\rightarrow - \phi$. The change of the sign of the transfer quasi-momentum ${\bf q}\rightarrow -{\bf q}$ corresponds to the change $\alpha\rightarrow \alpha+ \pi$.  

Let us write down explicit expression for the components of current density operators Eqs. (\ref{jPartial}) projected on the conduction band. We consider the geometry as given in the case {\bf (i)} above. 
\begin{eqnarray}
	&& {\bf j}_K({\bf p})=-ta\left\{ \left(\begin{array}{c}
		-1 \\ 0
	\end{array}\right)+ 
	\left(\begin{array}{c}
		1/2 \\ \sqrt{3}/2
	\end{array}\right) e^{i\frac{2\pi}{3}} e^{i\sqrt{3}a p_F\cos(\alpha+\phi-\pi/6)} 
	+  	\left(\begin{array}{c}
		1/2 \\ -\sqrt{3}/2
	\end{array}\right) e^{-i\frac{2\pi}{3}} e^{i\sqrt{3}a p_F\cos(\alpha+\phi+\pi/6)}  \right\} \label{jK_p} \\ 
	\nonumber 	&& {\bf j}_{K'}({\bf p}+{\bf q})={\bf j}_{K'}({\bf p'})=   -ta\left\{ \left(\begin{array}{c}
		-1 \\ 0
	\end{array}\right)+ 
	\left(\begin{array}{c}
		1/2 \\ \sqrt{3}/2
	\end{array}\right) e^{-i\frac{2\pi}{3}} e^{-i\sqrt{3}a p_F\cos(\alpha-\phi-\pi/6)} 
	+  	\left(\begin{array}{c}
		1/2 \\ -\sqrt{3}/2
	\end{array}\right) e^{i\frac{2\pi}{3}} e^{-i\sqrt{3}a p_F\cos(\alpha-\phi+\pi/6)}  \right\} \\ 
	\label{jK'_p'} 
\end{eqnarray}
The projection of the current density operator on the conduction band  is calculated using Eqs. (\ref{jK_p}), (\ref{jK'_p'}) as follows 
\begin{eqnarray}
	\nonumber 	&& 
	\langle {\bf p}+{\bf q}, K'|{\bf J}_{K'K}({\bf p}+{\bf q}, {\bf p})|{\bf p}, K\rangle_c= 
	\frac{1}{2} \left(e^{-i\phi'/2}, e^{i\phi'/2} \right) \left(\begin{array}{cc} 
		0 & {\bf j}_K({\bf p}) \\ 
		{\bf j}_{K'}^*({\bf p}+{\bf q}) & 0  
	\end{array} \right) 
	\left(\begin{array}{c}
		e^{-i\phi/2} \\  e^{i\phi/2} 
	\end{array}\right) = \\ 
	&& 
	\frac{1}{2}\left\{ {\bf j}_K({\bf p}) e^{i(\phi-\phi')/2}  + {\bf j}_{K'}^*({\bf p}+{\bf q}) e^{-i(\phi-\phi')/2} \right\}, 
	\label{JK'K_projected}
\end{eqnarray}
\begin{eqnarray}
	\nonumber 	&& 
	\langle {\bf p}, K|{\bf J}_{KK'}({\bf p}, {\bf p}+{\bf q} )|{\bf p}+{\bf q}, K'\rangle_c=  \langle {\bf p}+{\bf q}, K'|{\bf J}_{K'K}({\bf p}+{\bf q}, {\bf p})|{\bf p}, K\rangle_c^* = \\ 
	&& 
	\frac{1}{2}\left\{ {\bf j}^*_K({\bf p}) e^{-i(\phi-\phi')/2}  + {\bf j}_{K'}({\bf p}+{\bf q}) e^{i(\phi-\phi')/2} \right\}, 
	\label{JKK'_projected}
\end{eqnarray}
where $\phi=\phi_{\bf p}$ and $\phi'=\phi_{\bf p'}$ as defined by Eqs. (\ref{phiphi'_i}) , (\ref{phiphi'_ii}).  

For further calculations we also need the expressions for the projected interaction vertices 
\begin{eqnarray}
	&& \langle {\bf p+q}, K'| u_0\sigma_0| {\bf p},K\rangle_c= -\frac{1}{2} e^{-\frac{i}{2}|q| a \cos\alpha} \cos\left(\frac{|q| a}{2} \cos\alpha\right) \sin\alpha, \\ 
	&& \langle {\bf p+q}, K'| u_3\sigma_z| {\bf p},K\rangle_c= \frac{1}{2} e^{-\frac{i}{2}|q| a \cos\alpha} \sin\left(\frac{|q| a}{2} \cos\alpha\right) \cos\alpha , \\ 
	\nonumber 	&& \langle {\bf  p+q}, \! K'| u_1\sigma_x| {\bf p},\! K\rangle_c\!=\! \frac{U_1}{2 U_0} e^{-\frac{i}{2}|q| a \cos\alpha} \sin\phi \left\{1\!-\! \cos\left[ p_F a\sin\phi \cos\left(\frac{\pi}{3}\! -\! \alpha\right)\right]\! - \! \cos\left[ p_F a\sin\phi \cos\left(\frac{2\pi}{3}\! - \! \alpha\right)\right]  \right\}, \\ 
	&& \\ 
	\nonumber 	&& \langle {\bf p+q}, K'| u_2\sigma_y| {\bf p},K\rangle_c= -\frac{U_1}{2 U_0} e^{-\frac{i}{2}|q| a \cos\alpha} \cos\phi \left\{\sin\left[ p_F a\sin\phi \cos\left(\frac{\pi}{3}\!-\!\alpha\right)\right]\! + \!  \sin\left[ p_F a\sin\phi \cos\left(\frac{2\pi}{3}\! -\! \alpha\right)\right]  \right\}. \\ 
	&& 
\end{eqnarray}

\section{Perturbative calculation of the AC drag current}

\begin{figure}[t!]
	\centering
	\includegraphics[width=0.5\textwidth]{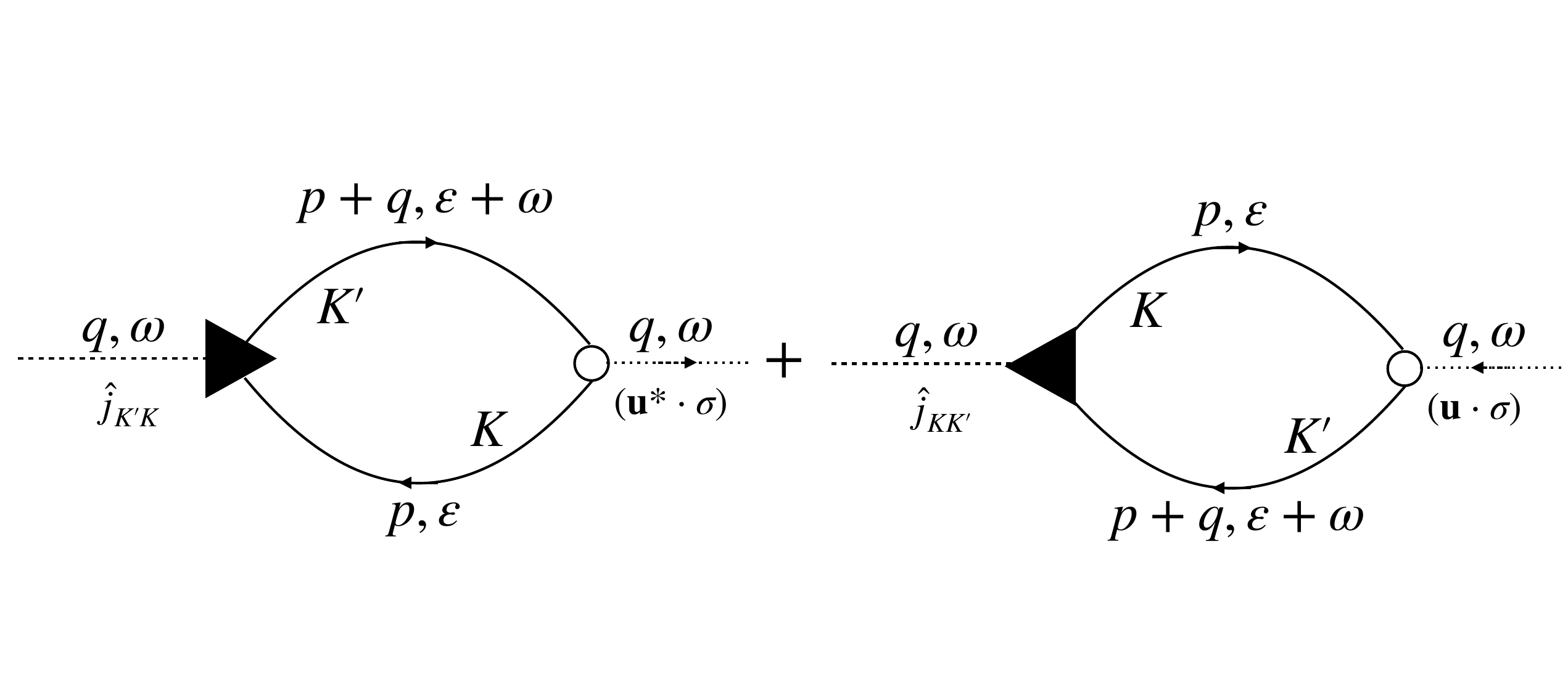}
	\caption{Diagrams for the lowest order contribution to the AC current.}
	\label{fig:DiagCurrent}
\end{figure} 

The diagram in Fig. \ref{fig:DiagCurrent} corresponds to the lowest order contribution to the current density due to the intervalley scattering $(KK')$. The mathematical expression reads 
\begin{equation}
	\langle {\bf j}_{K'K}({\bf q}, \omega)\rangle =-i U_0 \mathrm{Tr}\left\{ \gamma^q {\bf J}_{K'K}( {\bf p}+{\bf q}, {\bf p}) {\hat G}_{K}({\bf p}, \epsilon) \left( {\bf u}^*({\bf q, p})\cdot {\bm \sigma}\right) \gamma^{\mathrm{cl}}   {\hat G}_{K'}({\bf p}+{\bf q}, \epsilon+\omega) \right\}. 
\end{equation}
Here $\gamma^q=\sigma_1$ and $\gamma^{\mathrm{cl}}=\sigma_0$ denote the Pauli matrices acting in the Keldysh (RA) space. 

Using the notations introduced in Eq. (\ref{NotationsGF}),  we separate the parts reflecting the pole structure of Green functions (the spectrum) and the part corresponding to the matrix elements of the transition in 
the pseudo-spin and valley spaces 
\begin{eqnarray}
	\nonumber && 
	\langle {\bf j}_{K'K}({\bf q}, \omega)\rangle =-i U_0 \int_{\epsilon, {\bf p} } I_{K'K} \left\{ G^{K}({\bf p}, \epsilon) G^A({\bf p}+{\bf q}, \epsilon+\omega) + G^{R}({\bf p}, \epsilon) G^K({\bf p}+{\bf q}, \epsilon+\omega) \right\}= \\ 
	\nonumber && 
	-i U_0 \int_{\epsilon, {\bf p} } I_{K'K} \left\{ \tanh\left(\frac{\epsilon-\mu}{2T}\right) \left[ G^{R}({\bf p}, \epsilon)-G^{A}({\bf p}, \epsilon) \right]  G^A({\bf p}+{\bf q}, \epsilon+\omega)  +  \right. \\ 
	\nonumber && 
	\left. 
	G^{R}({\bf p}, \epsilon) \left[G^R({\bf p}+{\bf q}, \epsilon+\omega) 
	- G^A({\bf p}+{\bf q}, \epsilon+\omega) \right] 
	\tanh\left(\frac{\epsilon+\omega-\mu}{2T}\right)  \right\}= \\
	&& 
	-i U_0 \int_{\epsilon, {\bf p} } I_{K'K} \left\{ \left[\tanh\left(\frac{\epsilon-\mu}{2T}\right)- \tanh\left(\frac{\epsilon+\omega-\mu}{2T}\right)  \right]  G^{R}({\bf p}, \epsilon) G^A({\bf p}+{\bf q}, \epsilon+\omega) 
	\right\}, 
\end{eqnarray}
where in the last line we omitted the products of two retarded and two advanced Green functions, and introduced the notation 
\begin{equation}
	I_{K'K}(\epsilon, {\bf p}; \omega, {\bf q})=\mathrm{tr} \left\{ {\bf J}_{K'K}({\bf p}, {\bf p}+{\bf q})   g_{K}({\bf p}, \epsilon)  \left( {\bf u}^*({\bf q, p})\cdot {\bm \sigma}\right)  g_{K'}({\bf p}+{\bf q}, \epsilon+\omega)  \right\} 
	\label{IK'K}
\end{equation} 
Analogously, the contribution to the current by the scattering from the valley $K'$ to the valley $K$ is given by 
\begin{eqnarray}
	&& 
	\langle {\bf j}_{KK'}({\bf q}, \omega)\rangle =
	-i U_0 \int_{\epsilon, {\bf p} } I_{KK'} \left\{ \left[\tanh\left(\frac{\epsilon+\omega-\mu}{2T}\right)- \tanh\left(\frac{\epsilon-\mu}{2T}\right)  \right]  G^{R}({\bf p+q}, \epsilon+\omega) G^A({\bf p}, \epsilon) 
	\right\}, 
\end{eqnarray}
where 
\begin{equation}
	I_{KK'}(\epsilon, {\bf p}; \omega, {\bf q})=\mathrm{tr} \left\{ {\bf J}_{KK'}({\bf p}, {\bf p}+{\bf q}) g_{K'}({\bf p+q}, \epsilon+\omega)  \left( {\bf u}({\bf q, p})\cdot {\bm \sigma}\right)  g_{K}({\bf p}, \epsilon)  \right\}.
	\label{IKK'}
\end{equation}

\subsection{Momentum and frequency integration}
Closing the integration contour in the lower $\epsilon$ half-plane, we reduce the integral to the sum over the residue of the poles of the retarded Green function, which results in 
\begin{eqnarray}
	\nonumber && 
	\langle {\bf j}_{K'K}({\bf q}, \omega)\rangle \frac{-\pi  U_0}{2} \int_{\epsilon, {\bf p}}  I_{K'K} 	\left[\tanh\left(\frac{\epsilon+\omega-\mu}{2T}\right)- \tanh\left(\frac{\epsilon-\mu}{2T}\right)  \right] \times \\  
	\nonumber && 
	\left\{
	\delta(\epsilon-v_F|p|)\left[\frac{1}{\omega+v_F|p|-v_F|p+q|-i/\tau} + \frac{1}{\omega+v_F|p|+v_F|p+q|-i/\tau} \right]+ \right. \\ 
	&& \left. 
	\delta(\epsilon+v_F|p|)\left[\frac{1}{\omega-v_F|p|-v_F|p+q|-i/\tau} + \frac{1}{\omega-v_F|p|+v_F|p+q|-i/\tau} \right]
	\right\}
	\label{JK'K_momentum}
\end{eqnarray}

\begin{eqnarray}
	\nonumber && 
	\langle {\bf j}_{KK'}({\bf q}, \omega)\rangle = \frac{-\pi  U_0 }{2} \int_{\epsilon, {\bf p}}  I_{KK'} 	\left[\tanh\left(\frac{\epsilon+\omega-\mu}{2T}\right)- \tanh\left(\frac{\epsilon-\mu}{2T}\right)  \right] \times \\  
	\nonumber && 
	\left\{ 
	\delta(\epsilon+\omega-v_F|{\bf p+q}|)\left[\frac{1}{-\omega+v_F|{\bf p+q}|-v_F|p|-i/\tau} + \frac{1}{-\omega+v_F|{\bf p+q}|+v_F|p|-i/\tau} \right]+ \right. \\ 
	&& \left. 
	\delta(\epsilon+\omega+v_F|{\bf p+ q} |)\left[\frac{1}{-\omega-v_F|{\bf p+q}|-v_F|p|-i/\tau} + \frac{1}{-\omega-v_F|{\bf p+q}|+v_F|p|-i/\tau} \right]
	\right\} 
	\label{JKK'_momentum}
\end{eqnarray}
Here we introduced  small imaginary part at the poles $io\rightarrow \frac{i}{2\tau}$, which accounts phenomenologically for intrinsic  scattering processes in graphene.   

Furthermore, leaving only the pole contributions to the integrals over momentum ${\bf p}$, we obtain 
\begin{eqnarray}
	\nonumber && 
	\langle {\bf j}_{K'K}({\bf q}, \omega)\rangle = \frac{i \pi^2 U_0}{2} \int_{{\bf p}}\left\{ I_{K'K} (\epsilon=v_F |p|-\omega) 	
	\left[\tanh\left( \frac{\omega+ v_F |p|-\mu}{2T}
	\right)- \tanh\left(\frac{v_F |p| -\mu}{2T}\right)  \right] \times \right. \\  
	\nonumber && 
	\left[\delta(\omega+v_F|p|-v_F|p+q|) + \delta(\omega+v_F|p|+v_F|p+q|)\right] +\\ 
	\nonumber && 
	I_{K'K} (\epsilon=-v_F |p|) 	\left[\tanh\left(\frac{\omega- v_F |p|-\mu}{2T}\right)- \tanh\left(\frac{-v_F |p| -\mu}{2T}\right)  \right]\times \\ 
	\nonumber 	&& \left. 
	\left[\delta(\omega-v_F|p|-v_F|p+q|) + \delta(\omega-v_F|p|+v_F|p+q|) \right]
	\right\} \approx \\ 
	&&
	\frac{i \pi^2 U_0}{2} \int_{\bf p} I_{K'K} (\epsilon=v_F |p|) 	\left[\tanh\left(\frac{\omega+ v_F |p|-\mu}{2T}\right)- \tanh\left(\frac{v_F |p| -\mu}{2T}\right)  \right]  \delta(\omega+v_F|p|-v_F|p+q|),  \label{JK'K_int}
\end{eqnarray}

\begin{eqnarray}
	\nonumber && 
	\langle {\bf j}_{KK'}({\bf q}, \omega)\rangle = \frac{i \pi^2 U_0 }{2} \int_{{\bf p}}\left\{ I_{KK'} (\epsilon=v_F |{\bf p+q}|-\omega) 	\left[\tanh\left(\frac{v_F |{\bf p+q}|-\mu}{2T}\right)- \tanh\left(\frac{v_F |{\bf p+q}| -\omega -\mu}{2T}\right)  \right] \times \right. \\  
	\nonumber && 
	\left[\delta(v_F|{\bf p+q}|-\omega-v_F|p|) + \delta(-\omega+v_F|p|+v_F|{\bf p+q}|)\right] +\\ 
	\nonumber && 
	I_{KK'} (\epsilon=-\omega-v_F |{\bf p+q}|) 	\left[\tanh\left(\frac{- v_F |{\bf p+q}|-\mu}{2T}\right)- \tanh\left(\frac{-v_F |{\bf p+q}|-\omega  -\mu}{2T}\right)  \right]\times \\ 
	\nonumber && \left. 
	\left[\delta(-\omega-v_F|{\bf p+q}|-v_F|p|) + \delta(-\omega-v_F|{\bf p+q}|+v_F|p|) \right]
	\right\} \approx \\
	&& 
	\frac{i \pi^2 U_0 }{2} \int_{\bf p} I_{KK'} (\epsilon=v_F|p|) \left[\tanh\left(\frac{v_F |p|+\omega-\mu}{2T}\right)- \tanh\left(\frac{v_F|p|-\mu}{2T}\right)  \right] \delta(v_F|{\bf p+q}|-\omega-v_F|p|). 
	\label{JKK'_int}
\end{eqnarray}

In the last lines of Eqs. (\ref{JK'K_int}), (\ref{JKK'_int}) we left  the leading contributions at positive chemical potential and low temperatures. The combination of $\tanh$-functions in that term together with the energy and quasi-momentum conservation imposed by the $\delta$-function projects the integrals on the states of conduction band. 

It is convenient to perform the two-dimensional integration over momenta using polar coordinates, $p=(|{\bf p}|, \phi)$ where $\phi$ is defined as the angle between  vectors ${\bf p}$ and ${\bf q}$. 
Furthermore, we satisfy the $\delta$-functions by performing the angular integration. The angular integrals are calculated according to the formula 
\begin{equation}
	\int d\phi \delta(f(\phi)) F(\phi)=\frac{F(\phi_0)}{|f'(\phi_0)|}, 
\end{equation}
where $f'(\phi_0)=\frac{df(\phi)}{d\phi}\vert_{\phi=\phi_0}$, and $\phi_0$ satisfies the relation $f(\phi_0)=0$.
Here we show explicitly the computation of the integral with $\delta(\omega+v_F|p|-v_F|p+q|)$ in Eq. (\ref{JK'K_int}). The angular dependence in the $\delta$-function is contained in the term 
\begin{equation}
	|p+q|=\sqrt{p^2+q^2+2pq\cos\phi}, 
\end{equation}
therefore 
\begin{equation}
	f(\phi)=\omega+v_F p - v_F\sqrt{p^2+q^2+2pq\cos\phi}, \, \, \, \, 
	f'(\phi)=\frac{v_F pq\sin\phi}{\sqrt{p^2+q^2+2pq\cos\phi}}=\frac{v_F pq\sin\phi}{|p+q|}= 
	\frac{v_F^2 pq\sin\phi}{\omega+v_F p}, 
\end{equation}
where in the last equation we used the condition imposed by the $\delta$-function. 	

Furthermore, in the physically relevant regime $v_F p\gg \omega$, we can simplify  
\begin{equation}
	f'(\phi)\approx v_F q\sin\phi. 
\end{equation}
At low temperature, $T\ll \omega$, the $\tanh$ can be replaced by a step function. Then, the integration over the absolute value of momentum $|p|$ can be performed by replacing $|p|\approx p_F$, whereas the difference of the step-functions  determines the integration range $\omega/v_F$. Thus, the low-temperature approximation to the integration can be summarized by the expression 
\begin{equation}
	\int_{\bf  p}  \left[\tanh\left(\frac{\omega+ v_F |p|-\mu}{2T}\right)- \tanh\left(\frac{v_F |p| -\mu}{2T}\right)  \right] \delta(v_F|{\bf p+q}|-\omega-v_F|p|) ... \approx 
	\frac{\omega p_F}{v_F^2 |q|}  \int d|p| \int \frac{d\phi}{\sin\phi} \delta(|p|-p_F)\delta(\phi-\phi_0) ... , 
\end{equation}
where $\phi_0$  is the angle between $\bf p$ and $\bf q$ that solves the equation  
$|{\bf p+q|}=|p|=p_F$, which is  explicitly  given by 
\begin{equation}
	\phi_0=\arccos \left(-\frac{q}{2p_F}\right). 
	\label{phi_0}
\end{equation}

Finally, the AC drag current density, given by the sum of Eqs. (\ref{JK'K_int}) and  (\ref{JKK'_int})  becomes 
\begin{eqnarray}
	\nonumber && 
	{\bf j}_{\mathrm{AC}}= \frac{i \pi^2 U_0}{2} \int_{{\bf p}} \left( I_{K'K} (|p|, {\bf q}, \phi) +I_{KK'} (|p|, {\bf q}, \phi) \right) 	\left[\tanh\left(\frac{\omega+ v_F |p|-\mu}{2T}\right)- \tanh\left(\frac{v_F |p| -\mu}{2T}\right)  \right]
	\delta(\omega+v_F|p|-v_F|p+q|) = \\ 
	\nonumber &&
	\frac{i \pi^2 U_0}{2v_F |q| \sin\phi_0}  
	\int p dp \left( I_{K'K} (|p|, {\bf q}, \phi_0)  	+ I_{KK'} (|p|, {\bf q}, \phi_0) \right)  
	\left[\tanh\left(\frac{\omega+ v_F |p|-\mu}{2T}\right) - 
	\tanh\left(\frac{v_F |p| -\mu}{2T}\right)  \right] \approx \\ 
	&& 
	\frac{i \pi^2 U_0}{2v_F |q| \sin\phi_0}\frac{p_F\omega}{v_F} \left(  I_{K'K} (p_F, {\bf q}, \phi_0) +I_{KK'} (p_F, {\bf q}, \phi_0)\right)  = 
	\frac{i \pi^2 U_0 \omega p_F}{2v_F |q|  |\sin\phi_0|}  \left( \mathcal{I}_{K'K} (p_F, {\bf q}, \phi_0) + \mathcal{I}_{K'K} (p_F, {\bf q}, \phi_0) \right),  
	\label{JAC}
\end{eqnarray}
where in the last line  we introduced the dimensionless partial current densities  $\mathcal{I}_{ab}$ defined by the relation 
\begin{equation}
	I_{ab} (p_F, {\bf q}, \phi_0)	= v_F  \mathcal{I}_{ab} (p_F, {\bf q}, \phi_0). 
\end{equation}

The drag current given by Eq. (\ref{JAC}) formally diverges at the threshold $p_F=\frac{1}{2}(q-\omega/v_F)$, which corresponds to the angle $\phi_0=\pi$.   That divergence is removed  if one takes into account the scattering processes that change momentum of quasipartcles. Therefore, close to the threshold, one should take into account the finite life-time $\tau$ in Eqs.  (\ref{JK'K_momentum}), 
(\ref{JKK'_momentum})  explicitly, which amounts to the replacement $\sin\phi_0\rightarrow \frac{1}{\sqrt{v_F q\tau}}$ in Eq. (\ref{JAC})  at the cut-off $\phi_0=\pi$. The exact dependence of the prefactor in Eq. (\ref{JAC})  on the angle $\phi$ exhibits a crossover from the constant $ \sqrt{v_F q\tau}$ for the angles satisfying $\sin\phi\ll\frac{1}{\sqrt{v_F q\tau}}$ to $1/\sin\phi$ for $\sin\phi\gg \frac{1}{\sqrt{v_F q\tau}}$. 

\subsection{Drag current density}
In terms of the matrix elements projected onto the conduction band, the drag current density given by Eq. (\ref{JAC})  can be cast in the form 
\begin{eqnarray}
	\nonumber 
	{\bf j}_{\mathrm{AC}} \approx  \frac{i \pi^2 U_0 \omega p_F}{2v_F |q| |\sin\phi|} \left\{ \langle {\bf p}+{\bf q}, K'|{\bf J}_{K'K} ({\bf p}+{\bf q}, {\bf p}) 
	|K, {\bf p}\rangle\langle {\bf p}, K|{\bf u}^* {\bm{\sigma}} |{\bf p}+{\bf q}, K'\rangle +   
	\right. \\ 
	\nonumber 	 \left. 
	\langle {\bf p}, K|{\bf J}_{KK'} ({\bf p}, {\bf p}+{\bf q}) 
	|K', {\bf p}+{\bf q}\rangle \langle {\bf p}+{\bf q} , K'|{\bf u}{\bm{\sigma}} |{\bf p}, K\rangle 
	\right\}   \\ 
	\nonumber 
	= \frac{i \pi^2 U_0 \omega p_F}{2v_F |q| |\sin\phi|} 
	\Re\left[ \left({\bf j}_K({\bf p}) e^{i(\phi-\phi')/2}  + {\bf j}_{K'}^*({\bf p}+{\bf q}) e^{-i(\phi-\phi')/2}\right)\langle {\bf p}, K|{\bf u}^* {\bm{\sigma}} |{\bf p}+{\bf q}, K'\rangle  \right] 	\\ 
	=\frac{i \pi^2 U_0 \omega p_F}{2v_F |q||\sin\phi|} {\bf F}^{\mathrm{AC}}_{K'K}(\alpha, \phi) 
	\label{J_ac}
\end{eqnarray}
where the dimensionless vector function ${\bf F}^{\mathrm{AC}}_{K'K}(\alpha, \phi)$ is determined by the matrix elements of the current density operator and interaction vertices. It depends only on the directions of the transferred quasi-momentum and the momenta of the initial and final scattered states, parametrized by the angles $\alpha, \phi$ as shown in Fig. \ref{fig:Angles}.

\end{document}